\pdfoutput=1
\documentclass[letterpaper,twocolumn,10pt]{article}
\usepackage{usenix2019_v3}

\usepackage[normalem]{ulem}

\usepackage{times}

\usepackage{prp-macros}

\usepackage{inconsolata}

\usepackage{tikz}
\usepackage{pgfplots}
\usepackage[ruled, noend, linesnumbered, commentsnumbered]{algorithm2e}

\newcommand{\sys}{SGX-LKL\xspace}

\newcommand{\sgxlkllib}{\emph{libsgxlkl.so}\xspace}
\newcommand{\sysloader}{\emph{sgx-lkl-run}\xspace}

\usepackage{caption}
\usepackage{float}
\usepackage[small,compact]{titlesec}

\begin{document}

\title{\sys: Securing the Host OS Interface for Trusted Execution}

 \author{
 Christian Priebe\qquad Divya Muthukumaran\qquad Joshua Lind\qquad Huanzhou Zhu\\Shujie Cui\qquad Vasily A. Sartakov\qquad Peter Pietzuch\\{\small{}Imperial College London}
 } 

%

\date{}

\maketitle

\begin{abstract}
  Hardware support for trusted execution in modern CPUs enables tenants to
  shield their data processing workloads in otherwise untrusted cloud
  environments. Runtime systems for trusted execution must rely on an interface
  to the untrusted host OS to use external resources such as storage, network,
  and other functions. Attackers may exploit this interface to leak data or
  corrupt the computation.
 
  We describe \emph{\sys}, a system for running Linux binaries inside of Intel
  SGX enclaves that exposes only a minimal, protected and oblivious host
  interface: the interface is (i)~\emph{minimal} because \sys uses a complete
  library OS inside the enclave, including file system and network stacks,
  leading to an interface with only 7~calls; (ii)~\emph{protected} because \sys
  transparently encrypts and integrity-protects all data passed via low-level
  I/O operations; and (iii)~\emph{oblivious} because \sys performs host
  operations independently of the application workload: for oblivious disk I/O,
  it uses an encrypted ext4 file system with randomised disk blocks; for
  oblivious network I/O, it uses layer-3 encryption with dummy packets for
  constant bit rates between trusted nodes. We show that \sys can protect
  TensorFlow training with only a 21\% overhead.
\end{abstract}



\section{Introduction}
\label{sec:introduction}

In healthcare~\cite{oliver2006healthgear, al2012security, zhang2010security}
and finance~\cite{chellappa2002perceived, kim2008trust}, it is challenging to
run sensitive workloads in public cloud environments due to security
concerns~\cite{csacloudthreats}. For example, when training a machine learning
model, both the training data and the model may be confidential, and the
integrity of the model must be ensured.
In clouds, the data and computation may be disclosed or corrupted due to
infrastructure bugs~\cite{gunawi2014bugs},
misconfigurations~\cite{techrepublicdropboxleaks}, rogue
administrators~\cite{insiderAttacks, duncan2012insider}, or external
attacks~\cite{gruschka2010attack, CloudAttacks}.

Modern Intel~\cite{sgx14}, ARM~\cite{Trustzone} and AMD~\cite{AMDSEV} CPUs
offer hardware support for \emph{trusted execution
  environments}~(TEEs)~\cite{sabt2015trusted}. A TEE protects the
confidentiality and integrity of computation and data by shielding it from the
rest of the system. Intel's \emph{Software Guard Extensions}~(SGX)~\cite{sgx14}
add new instructions to execute code within isolated memory regions called
\emph{enclaves}. Enclaves are encrypted by the CPU.

Cloud providers have begun to roll out support for
TEEs~\cite{microsoftconfidential, ibmcloud}. This makes a ``lift-and-shift''
model---in which tenants move a whole application to a TEE---an attractive
proposition. TEE runtime systems, such as Haven~\cite{baumann2014haven},
SCONE~\cite{arnautov2016scone} and Graphene-SGX~\cite{tsai2017graphene}, have
demonstrated the feasibility of executing complete Linux applications inside
TEEs with acceptable performance overheads.

When applications inside a TEE require external resources, such as files, the
network or other OS functions, they must rely on the untrusted host OS. TEE
runtime systems therefore have a host interface, but its security implications
are poorly understood and handled: the host interface may (i)~accidentally
expose state from the TEE to the outside by leaking sensitive data in
calls~\cite{ta2006splitting}; (ii)~act as a side channel where the existence or
absence of a call reveals application state~\cite{wang2018interface}; and
(iii)~have a malicious implementation and thus compromise application integrity
inside the TEE~\cite{iago_attack}.

Existing TEE runtime systems vary in the types of interfaces that they expose
to hosts, and the nature of the interface is a consequence of their design.
%
%
We explore the opposite approach: we begin our design of a TEE runtime system
with a desired secure host interface, and then decide on the necessary system
support inside the TEE. We aim for a host interface with three properties:
(1)~\emph{minimality}---only functionality that cannot be provided inside the
TEE should be part of it; (2)~\emph{protection}---all data that crosses the
host interface must be encrypted and integrity-protected; and
(3)~\emph{obliviousness}---the presence or absence of host calls should not
disclose information about the application state.

We describe \textbf{\sys}, a new TEE runtime system that executes unmodified
Linux binaries inside SGX enclaves while exposing a minimal, data-protected
and oblivious host interface.\footnote{\sys is available as open-source
software: \url{https://github.com/lsds/sgx-lkl}.} The design of \sys makes
three contributions:


\tinyskip

\noindent
\textbf{(1)~Minimal host interface~(\S\ref{sec:design}).} \sys only requires
7~host interface calls, which expose low-level functionality for block-level
I/O for storage and packet-level I/O for networking. Higher-level POSIX
functionality is implemented in \sys by porting a complete Linux-based library
OS to an SGX enclave. \sys uses the \emph{Linux Kernel
  Library}~(LKL)~\cite{purdila2010lkl} for a mature POSIX implementation with a
virtual file system layer and a TCP/IP network stack.

\tinyskip

\noindent
\textbf{(2)~Protected host interface~(\S\ref{sec:hardening}).} \sys ensures
that all I/O operations across the host interface are encrypted and
integrity-protected transparently: (i)~for file I/O, \sys uses an encrypted
Linux \emph{ext4} root file system image stored outside of the SGX enclave,
which is accessed using the Linux \emph{device mapper}
subsystem~\cite{devicemapper}; and (ii)~for network I/O, \sys creates a
\emph{virtual private network}~(VPN) overlay that secures all network
traffic. Layer-3 IP packets are encrypted by the in-kernel
Wireguard~\cite{wireguard} VPN implementation.

To verify the integrity of a \sys instance and provide cryptographic keys, \sys
supports a runtime attestation and provisioning process: it first attests the
integrity of the \sys implementation inside the SGX enclave and then provisions
the instance with the keys required to access the encrypted root file system
image and the VPN channels.

\tinyskip

\noindent
\textbf{(3)~Oblivious host interface~(\S\ref{sec:hardening:oblivious}).} To
prevent any information leakage as part of the host interface, \sys makes the
calls independent of the application workload inside of the SGX enclave:
(i)~it executes host calls in fixed batches, and each batch includes the same
number of calls by adding indistinguishable dummy calls; (ii)~for block I/O
requests, \sys uses an oblivious construction that reuses the existing
\emph{ext4} file system image format. The encrypted blocks in the file system
image are randomised and shuffled, which only exposes random access patterns
to the host. Between shuffles, \sys reads a block at most once by relying on
the in-enclave page cache; and (iii)~for network I/O, \sys sends fixed-size
packets and makes use of Linux' traffic shaping capabilities to ensure a
constant traffic bitrate between trusted nodes.

\tinyskip

\noindent
Our experimental evaluation shows that \sys{}'s host interface has a reasonable
performance overhead. With emulated SGX enclaves (to ignore current SGX memory
limitations) and oblivious calls, \sys trains common deep neural network models
using TensorFlow~\cite{tensorflow} with an overhead of 14\%--21\% on one
machine and 2.7$\times$--2.9$\times$ in a distributed deployment. With SGX
hardware, \sys runs PARSEC benchmarks that fit into SGX memory with an overhead
of 3.1$\times$ and 1.5$\times$ with and without oblivious calls, respectively.



\section{Host Interfaces for Trusted Execution}
\label{sec:background}


\subsection{Trusted execution environments in clouds}
\label{sec:background:trusted_execution}

We assume that a cloud provider supports CPU-implemented \emph{trusted
  execution environments~(TEEs)}. TEEs separate userspace code from the rest of
the system, including privileged software such as the
OS~\cite{tee_model}. Multiple TEE implementations are available, \eg Intel
SGX~\cite{sgx14}, ARM TrustZone~\cite{Trustzone} and AMD SEV~\cite{AMDSEV},
with others under way~\cite{lee2019keystone, multizone, optee}.

Intel's \emph{Software Guard Extensions}~(SGX)~\cite{sgx14} provide new CPU
instructions to create TEEs called \emph{enclaves}. Enclave memory is encrypted
and integrity-protected transparently by the CPU. The CPU controls transitions
from untrusted to enclave code, and only enclave code can access enclave
memory. Due to this isolation, enclave code has no I/O access but must use the
untrusted host.

Intel SGX also supports \emph{attestation}, allowing a remote party to validate
the enclave code. For this, the CPU (i)~measures the enclave's contents by
computing a hash; (ii)~signs the measurement hash; and (iii)~provides it to the
attesting party. The attestor can verify the signature and the hash.

\label{sec:background:app_model}

For the deployment of applications using TEEs, we focus on a
\emph{lift-and-shift} model~\cite{lift-shift}: users deploy unmodified Linux
binaries. For binary compatibility, this requires a \emph{TEE runtime
  system}~\cite{baumann2014haven, arnautov2016scone,tsai2017graphene}, which
provides POSIX abstractions. TEE runtime systems must use functionality by the
untrusted host OS for operations outside of the trust domain of the TEE, \eg
when using I/O resources. We refer to the interface between the TEE runtime
system and the host as the \emph{host interface}.

\subsection{Security goals and threat model}
\label{sec:background:threat_model}

We want to prevent an adversary from compromising the \emph{confidentiality}
and \emph{integrity} of the application code and its input and output data.
%
%
We consider all software outside the TEE, including the host OS kernel, as
under adversarial control.

We assume that the TEE itself is trustworthy.
%
%
Existing attacks against TEEs, \eg Spectre~\cite{lsds_spectre},
Foreshadow~\cite{van2018foreshadow}, Zombieload~\cite{schwarz2019zombieload},
and controlled-channel attacks~\cite{xu2015controlled} are orthogonal to our
work. They exploit flaws in current TEE implementations and can be mitigated
through hardware and/or microcode changes~\cite{microcode_updates}.  As the
maturity of TEE implementations, especially new open-source
ones~\cite{IntelSGX,AMDSEV,lee2019keystone}, grows over time, such attacks will
become rarer.
Cache side-channel attacks~\cite{MoghimiIE17, BrasserMDKCS17, GotzfriedESM17}
are not specific to TEEs and are enabled by micro-architectural resource
sharing. They require fundamental mitigations through compiler
techniques~\cite{Varys, DR.SGX, Hyperspace} and defensive
programming~\cite{osvik2006cache, sasy2018zerotrace, cui2018preserving}.

Our threat model instead focuses on attacks against the host interface because
they are easy to carry out and do not assume particular micro-architectural
behaviour. An adversary may compromise (i)~\emph{confidentiality}: they can
observe the parameters, frequencies and sequences of host calls. For example,
they can learn the disk I/O access pattern to determine the application or
workload being run---linear scans and repeated accesses are visible via the
host interface. This side channel discloses information about the TEE
execution; and compromise (ii)~\emph{integrity}: an adversary can modify the
input/output parameters of host calls, repeatedly perform arbitrary host calls
or interfere with their outside execution. For example, an adversary may trick
an enclave into exposing confidential data via \emph{Iago}
attacks~\cite{iago_attack} or hijack its control flow.


\renewcommand{\tabcolsep}{3pt}
\begin{table*}[tbp]
\scriptsize\linespread{0.75}\selectfont\centering
\caption{Security breakdown of parameters in host interface calls for existing
  TEE runtime systems}
\label{tbl:TEEinterfaces}
\begin{tabular}{lll ccccc|ccccc}
\toprule
 \multicolumn{1}{l}{\multirow{4}{*}{
 	\textbf{\begin{tabular}[c]{@{}l@{}}TEE\\runtime\\system\end{tabular}}}}
 	&\multicolumn{1}{l}{\multirow{4}{*}{
		\textbf{\begin{tabular}[c]{@{}c@{}}Function\end{tabular}}}}
 	&\multicolumn{1}{l}{\multirow{4}{*}{
 		\textbf{\begin{tabular}[c]{@{}l@{}}Number of\\ host calls
                         \end{tabular}}}}
	&\multicolumn{5}{c}{\begin{tabular}[c]{@{}c@{}}\textbf{Out parameters (impact confidentiality)} \\ More difficult to protect \\ $\xrightarrow{\hspace*{4cm}}$\end{tabular}}
	&\multicolumn{5}{c}{\begin{tabular}[c]{@{}c@{}}\textbf{In parameters (impact integrity)} \\ More difficult to protect \\ $\xrightarrow{\hspace*{4cm}}$ \end{tabular}}\\ \cmidrule(l{2pt}r{2pt}){4-8} \cmidrule(l{2pt}r{2pt}){9-13}
\multicolumn{1}{l}{}
	&\multicolumn{1}{c}{}
	&\multicolumn{1}{c}{}
	&\multicolumn{1}{c}{{\begin{tabular}[c]{@{}c@{}}Variable\\ size buffer\end{tabular}}}
	&\multicolumn{1}{c}{{\begin{tabular}[c]{@{}c@{}}Address\\ range \end{tabular}}}
	&\multicolumn{1}{c}{{\begin{tabular}[c]{@{}c@{}}Pure\\ identifier \end{tabular}}}
	&\multicolumn{1}{c}{{\begin{tabular}[c]{@{}c@{}}Impure\\ identifier \end{tabular}}}
	&\multicolumn{1}{c}{{\begin{tabular}[c]{@{}c@{}}Semantic \\parameters\end{tabular}}}
	&\multicolumn{1}{c}{{\begin{tabular}[c]{@{}c@{}}Variable\\ size buffer\end{tabular}}}
	&\multicolumn{1}{c}{{\begin{tabular}[c]{@{}c@{}}Address\\ range \end{tabular}}}
	&\multicolumn{1}{c}{{\begin{tabular}[c]{@{}c@{}}Pure\\ identifier \end{tabular}}}
	&\multicolumn{1}{c}{{\begin{tabular}[c]{@{}c@{}}Impure\\ identifier \end{tabular}}}
	&\multicolumn{1}{c}{{\begin{tabular}[c]{@{}c@{}}Semantic \\parameters\end{tabular}}}
\\  \midrule
\multirow{4}{*}{
	\textbf{Panoply}~\cite{shinde2017panoply}}
	&{{I/O}}       &239  &11 &17 &96 &49 &139     &10  &2  &22  &24  &74
\\ 
	&{{Events}}    &22   &-- &-- &5  &-- &22      &--  &--  &2   &-- &8
\\ 

	&{{Time}}      &12   &-- &-- &-- &-- &7       &--  &--  &--  &-- &10
\\ 

	&{{Threading}} &29   &1  &-- &10 &-- &17      &2   &--  &4   &-- &10
\\ \midrule
	\multirow{4}{*}{\textbf{Graphene-SGX}~\cite{tsai2017graphene}}
	&{{I/O}}       &29   &3  &1  &18  &6 &19     &3  &2  &2  &4  &7
\\ 

	&{{Events}}    &1    &-- &-- &1  &-- &1      &-- &-- &-- &-- &1
\\ 

	&{{Time}}      &2    &-- &-- &-- &-- &1      &-- &-- &-- &-- &2
\\ 

	&{{Threading}} &6    &-- &--  &-- &1 &4      &1  &-- &1 &-- &1
\\ \midrule
	\multirow{4}{*}{\textbf{Haven}~\cite{baumann2014haven}}
	&{{I/O}}       &11   &1  &3  &6  &1  &7    &1  &-- &3  &-- &1
	\\ 

	&{{Events}}    &6    &-- &-- &5  &-- &1    &-- &-- &1  &-- &3
	\\ 

	&{{Time}}      &1    &-- &-- &-- &-- &--   &-- &-- &-- &-- &1
	\\ 

	&{{Threading}} &6    &-- &-- &1  &-- &2    &-- &-- &1  &-- &--
\\ \midrule\midrule
  \multirow{4}{*}{\textbf{\sys}}
	&{{I/O}}       &4   &2  &-- &-- &-- &2    &2  &-- &-- &-- &--
  \\ 

	&{{Events}}    &2   &-- &-- &-- &-- &1    &-- &1  &-- &-- &2
	\\ 

	&{{Time}}      &1   &-- &-- &-- &-- &--   &-- &-- &-- &-- &1
	\\ 

	&{{Threading}} &--  &-- &-- &-- &-- &--   &-- &-- &-- &-- &-- \\
\bottomrule
\end{tabular}
\end{table*}

\subsection{Host interfaces of current TEE runtime systems}
\label{sec:background:host_interfaces}

Next we analyse the security of the host interfaces of four existing TEE
runtime systems:

\tinyskip

\noindent
(1)~\emph{Panoply}~\cite{shinde2017panoply} minimises the size of the
\emph{trusted computing base}~(TCB) inside the TEE. It places the \emph{glibc}
standard C library~\cite{glibc} outside and implements a shim layer that
forwards calls. Panoply exposes the largest host interface because it delegates
all C library calls to the host.

\tinyskip

\noindent
(2)~\emph{SCONE}~\cite{arnautov2016scone} provides a modified \emph{musl}
standard C library~\cite{musl} alongside the application inside the
TEE. Compared to Panoply, SCONE has a larger TCB but a smaller host interface:
it forwards only system calls and not all libc calls to the host. It uses a
custom shim layer to protect I/O calls.

\tinyskip

\noindent
(3)~\emph{Graphene-SGX}~\cite{tsai2017graphene}
reduces the size of the host interface compared to Panoply and SCONE by
implementing a partial library OS inside the TEE. It relies on the host for
the file system, network stack, and threading implementations.

\tinyskip

\noindent
(4)~\emph{Haven}~\cite{baumann2014haven} uses the Drawbridge library
OS~\cite{Drawbridge} to execute Windows applications.
Drawbridge also provides file system and network stack implementations inside
the enclave.

\mypar{Types of host interface parameters} In our security analysis, we focus
on the parameters of host calls because they can leak information to the host
or compromise enclave integrity. We categorise parameters into five types,
ordered by the difficulty of protection from easiest to hardest:
(i)~\emph{variable-sized buffers} pass a user-defined byte array across the
host interface. They are used in file/network I/O operations (\eg the
\code{buf} and \code{count} parameters in the \code{read()}
call~\cite{readsystemcall});
(ii)~\emph{address ranges} represent parameters that refer to regions of
untrusted or trusted memory (\eg the parameters passed to \code{mmap()} and the
return value of \code{malloc()});
(iii)~\emph{pure/impure identifiers} point to entities:
%
%
pure identifiers only identify an entity (\eg a file descriptor) ; impure
identifiers also disclose information about the entity (\eg a path name); and
(iv)~\emph{semantic parameters} refer to parameters with opaque semantics
specific to the host call (\eg \code{mode} and \code{flags} for file access
operations).

\tinyskip

\noindent
\T\ref{tbl:TEEinterfaces} shows the number of host calls for each system, split
by function (I/O, events, time, threading).\footnote{For Panoply and
  Graphene-SGX, the host calls are taken from the GitHub source
  code~\cite{graphenesourcecode, panoplysourcecode}; for Haven, they are
  obtained from the paper~\cite{baumann2014haven}. We do not breakdown SCONE's
  interface because its implementation is not public. As SCONE forwards system
  calls, its host interface is similar to Panoply's.}  Panoply has a large
interface, with 302~calls in total; Graphene-SGX and Haven require 38~and
24~calls, respectively.


We break down the parameters according to the above types, distinguishing
between \emph{out} parameters, which are passed to the host and may compromise
confidentiality, and \emph{in} parameters, which are passed into the TEE and
may affect integrity. Return values of calls out of the enclave are considered
to be the same as \emph{in} parameters.


\mypar{Confidentiality attacks} To learn sensitive information, an adversary
may observe \emph{out} parameters:

\tinyskip

\noindent
(i)~\emph{Variable-sized buffers} may contain security-sensitive data. To
ensure confidentiality, such data must be encrypted and padded to a fixed
size. If the true size is exposed, an adversary may infer size-dependent
secrets. For example, in an image classification
application~\cite{KrizhevskySH17}, an adversary may learn the classification by
considering different result buffer lengths.

\T\ref{tbl:TEEinterfaces} shows that Panoply, Graphene-SGX and Haven use 11, 3
and 1 buffers, respectively, for I/O operations; Panoply also uses a buffer to
share messages between threads. Buffers are encrypted but their sizes are
disclosed. Haven has an in-enclave file system and writes data to the host disk
as fixed-sized blocks; SCONE divides files into fixed-size chunks for
authentication but reveals the number of chunks; Graphene-SGX and Panoply do
not provide transparent file encryption. SCONE, Panoply, and Graphene all
reveal file sizes.

\tinyskip

\noindent
(ii)~\emph{Address ranges} passed from the TEE to the host point to continuous
regions of untrusted memory. For I/O operations, Panoply, Graphene-SGX and
Haven expose 17, 1 and 3~address ranges, respectively. An adversary may observe
their usage pattern. For example, Panoply uses untrusted memory for
communication between enclaves that isolate application compartments. The usage
pattern reveals application-specific control flow, \eg secret-dependent
inter-enclave calls.

\tinyskip

\noindent (iii)~\emph{Pure/impure identifiers.} Panoply, SCONE, Graphene-SGX,
and Haven use a large number of identifiers for I/O operations, event handling
and threading. Similar to address ranges, passing identifiers to the host
reveals usage patterns. For example, Cash~\etal\cite{CashGPR15} show that an
adversary can learn database queries by observing record accesses, even if the
queries are executed within a TEE.

Generally, pure identifiers reveal less information than impure identifiers
(\eg file names), as pure identifiers can be chosen randomly. Preventing
information leakage by impure identifiers can only be done on a case-by-case
basis (\eg by replacing file names with hashes). The above systems do not
provide any protection for impure identifiers.

\tinyskip

\noindent
(iv)~\emph{Semantic parameters} have a context-specific meaning and therefore
cannot be encrypted transparently. Existing runtime systems
%
%
have a large number of unprotected semantic parameters. For example,
Graphene-SGX, Panoply, and Haven all rely on the host for thread
synchronisation. Graphene-SGX exposes address, operation, value and timeout
parameters of a \code{futex} host call. This gives an adversary detailed
information of the application state and enables attacks such as
\emph{AsyncShock}~\cite{weichbrodt2016asyncshock}, which exploit host control
of enclave threads.

\mypar{Integrity attacks} To compromise integrity, an adversary may tamper with
parameters. Integrity protecting \emph{in} parameters varies in difficulty
according to the parameter type:

\tinyskip

\noindent
(i)~\emph{Variable-sized buffers} passed to the TEE must have their integrity
protected. For I/O calls, Panoply, Graphene-SGX, and Haven use 10, 3 and 1
parameters with variable-sized buffers, respectively. Panoply and Graphene-SGX
also use variable-sized buffers for inter-enclave communication.

This exposes two means of integrity attacks: an adversary may modify (i)~the
buffer contents to violate data integrity; and (ii)~the buffer count to trigger
an overflow. To ensure buffer integrity, the contents must be protected by an
HMAC. The TEE must ensure the freshness of the HMAC, otherwise an adversary can
swap two valid buffers or perform a roll-back attack. Panoply protects the
integrity of the inter-enclave communication buffer using TLS but does not
protect file I/O; Graphene-SGX ensures the integrity of file I/O by maintaining
a Merkle tree of file chunk hashes inside the enclave. Neither Panoply,
Graphene-SGX nor Haven protect the integrity of network I/O for applications
without TLS support.

To prevent buffer overflow attacks, the TEE must check buffer lengths. All
analysed runtime systems do this.

\tinyskip

\noindent (ii)~\emph{Address ranges} are difficult to integrity check. For I/O
operations, Panoply and Graphene-SGX use 2~address ranges each. They check that
the ranges are fully inside or outside of TEE memory, preventing control flow
hijacking~\cite{BiondoCDFS18, lee2017hacking}.

An adversary, however, may still manipulate addresses by reusing old ranges to
roll back data, swapping ranges, or modifying ranges to corrupt data. The three
runtime systems are vulnerable to such attacks, and the only effective
mitigation is to avoid relying on untrusted address ranges.

\tinyskip

\noindent
(iii)~\emph{Pure/impure identifiers} are commonly used as \emph{in} parameters:
Panoply uses 46, 2, and 4 identifiers for I/O, events, and threading,
respectively; Graphene-SGX uses 6~identifiers for I/O and 1 for threading;
Haven uses 3, 1, and 1~identifiers for I/O, events and threading, respectively.

Adversaries may pass invalid or manipulated identifiers: if two file
descriptors are swapped, the enclave may access incorrect files. Such malicious
activity can be detected: Graphene-SGX maintains per-file HMACs, revealing
wrong file descriptors, and incorrect sockets are revealed by TLS.


\tinyskip

\noindent
(iv)~\emph{Semantic parameters} must have their integrity verified on a
case-by-case basis. All three systems use many semantic \emph{in} parameters
for I/O, events, time and threading operations.

For semantic parameters with few valid values, such as \code{errorcode} and
\code{signum}, the TEE can perform explicit checks; for ones with a larger
domain, such as \code{size}, bounds checks are possible. None of these checks
establish semantic correctness though.
Since both Panoply and Graphene-SGX rely on the host file system, they can only
check for the plausibility of returned file meta-data, such as
\texttt{st\_size} and \texttt{st\_blocks}.


\begin{table*}[tb]
\footnotesize\linespread{0.8}\centering
\caption{\sys{} host interface}
\label{tab:hostinterface}
\begin{tabular}{m{1.0cm} m{6.0cm} m{9.0cm}}
\toprule
& \textbf{Call} & \textbf{Description} \\
\midrule
\multirow{4}{*}{\textbf{I/O}}
& \texttt{disk\_read(offset, buf)} & Reads 4\unit{KB} block from disk image file at \texttt{offset} \\
& \texttt{disk\_write(offset, buf)} & Writes 4\unit{KB} block from \texttt{buf} to disk image file at \texttt{offset} \\
& \texttt{net\_read(buf)} & Reads MTU-sized Ethernet frame from network device into \texttt{buf} \\
& \texttt{net\_write(buf)} & Writes MTU-sized Ethernet frame from \texttt{buf} to network device \\
\midrule
\multirow{2}{*}{\textbf{Events}}
& \texttt{net\_poll(eventmask)} $\,\to\,$ \texttt{reventmask} & Waits for \texttt{eventmask} events on network device; returns occurred events \\
& \texttt{forward\_signal(num, code, addr)} & Forwards signal \texttt{num} with description \texttt{code} occurred at \texttt{addr} to enclave \\
\midrule
\textbf{Time} & \texttt{time\_read()} $\,\to\,$ \texttt{time} & Reads time from untrusted vDSO memory region \\
\bottomrule
\end{tabular}
\end{table*}


\subsection{Designing a secure host interface}
\label{sec:background:sgxlkl}

While host interfaces of existing runtime systems as well as SDKs have also
been shown to contain implementation flaws~\cite{vanbulck2019tale}, the above
issues related to the host interface are more fundamental. We conclude that it
is non-trivial (and at worst impossible) to protect \emph{out} parameters from
leaking information and exposing access patterns and to verify the correctness
of \emph{in} parameters. Therefore, the first step in designing a secure host
interface is to keep it narrow and minimise the number of parameters. Only
functionality that cannot be provided within an enclave should be delegated to
the host.

In addition, we want to reduce the number and complexity of host call
parameters. By building on low-level calls instead of high-level POSIX
abstractions, the parameter types become simpler. For example, Panoply, SCONE,
and Graphene-SGX rely on the host file system and thus must expose impure
identifiers such as file names. A secure host interface should avoid delegating
high-level resource management to the host.

Based on these principles, \T\ref{tbl:TEEinterfaces} shows the host calls and
parameter types of \emph{\sys}, our TEE runtime system. It avoids address range
and pure/impure identifiers and only requires a few simple semantic parameters
of which all but one (\code{time}) have a fixed set of valid values or can be
bounds-checked.



\section{Minimising the host interface}
\label{sec:design}

\T\ref{tab:hostinterface} shows \sys{}'s host interface, which has only 7~calls
with functionality that cannot be provided inside enclaves.


\subsection{\sys{} host interface}
\label{sec:design:host_interface}

\mypar{I/O operations} \sys uses a low-level I/O interface: for disk I/O,
\code{disk\_read()} and \code{disk\_write()} read/write a disk block from/to a
persistent block device, respectively. Each call takes an \code{offset} into
the block device and a pointer \code{buf} to a buffer; for network I/O,
\code{net\_read()} and \code{net\_write()} receive and send fixed-size network
packets using buffer \code{buf}.

\mypar{Events} A \code{net\_poll()} call passes an \code{eventmask} to the host
with the network events that \sys is waiting for. The call blocks until network
packets are available to be read or outgoing packets can be sent. It returns
which events have occurred.

\sys must handle hardware exceptions, such as page access violations or illegal
instructions. An exception causes an enclave exit and transfers control to the
host kernel. A \code{forward\_signal()} call provides the signal description to
the enclave: the signal number \code{num}, the cause \code{code}, and the
associated memory address \code{addr}. The exception is then either processed
by \sys directly or forwarded to the application if it has registered a
corresponding signal handler.

\mypar{Time} The call \code{time\_read()} reads time from different clock
sources. It is used by application code and by \sys, \eg to generate timer
interrupts required by LKL (see~\S\ref{sec:design:system_support}).

\subsection{In-enclave OS functionality}
\label{sec:design:system_support}

\begin{figure}[tb]
    \centering
    \includegraphics[width=.95\linewidth]{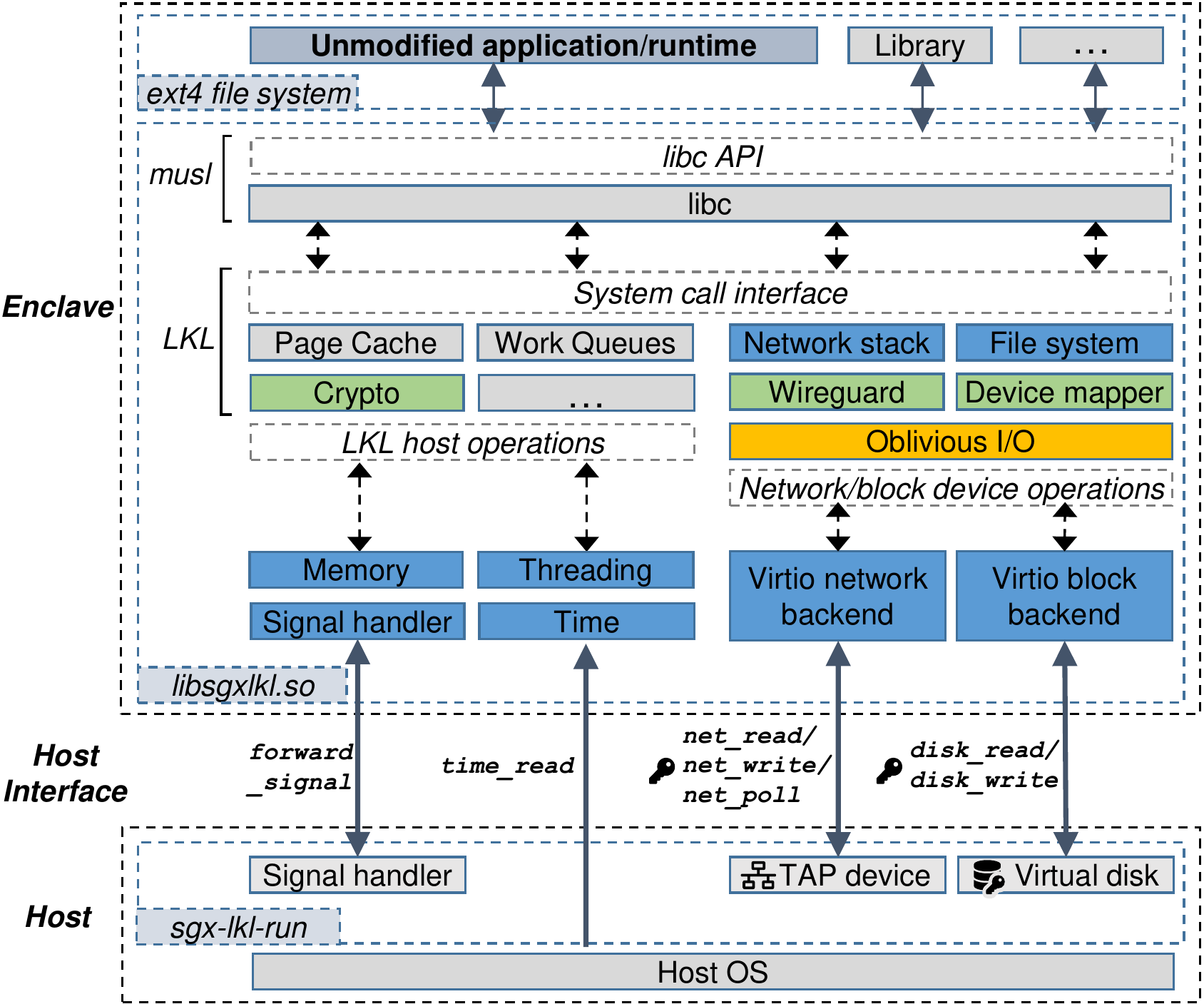}
    \caption{\sys{} architecture}
\label{fig:architecture}
\end{figure}

Its host interface allows \sys to access low-level resources, but Linux
applications require higher-level POSIX abstractions. To bridge this gap, \sys
provides the following OS functions inside the enclave: (i)~file system
implementations; (ii)~a TCP/IP network stack; (iii)~threading and scheduling;
(iv)~memory management; (v)~signal handling; and (vi)~time. Rather than
invoking these OS functions through systems calls directly, applications link
against a standard C library~(\emph{libc}). Similar to other
systems~\cite{arnautov2016scone, tsai2017graphene}, \sys includes a libc
implementation to support unmodified dynamically-linked binaries.

\F\ref{fig:architecture} shows the \sys architecture. Next we describe the OS
functionality provided inside the enclave in detail. Components covered in this
section are shown in blue; the parts responsible for protecting the host
interface (shown in green) and for making the interface oblivious (shown in
orange) are described in \S\ref{sec:hardening} and
\S\ref{sec:hardening:oblivious}, respectively.

An untrusted loader, \sysloader, creates the SGX enclave and loads the enclave
library \sgxlkllib, which runs alongside the application within the enclave.
\sgxlkllib includes a modified \emph{musl}~\cite{musl} C standard library that
redirects system calls to an in-enclave library OS provided by the \emph{Linux
  Kernel Library}~(LKL)~\cite{purdila2010lkl}. LKL is an architecture port of
the Linux kernel to userspace. It enables \sys to make components such as the
Linux kernel page cache, work queues, file system and network stack
implementations, and crypto libraries available inside the enclave. As it is
intended to run in userspace, LKL expects a set of \emph{LKL host operations},
\eg to create threads or to allocate memory. Therefore, \sgxlkllib includes
further components for memory management, user-level threading, signal
handling and time.

\mypar{File systems} Existing TEE runtime systems~\cite{arnautov2016scone,
  tsai2017graphene, shinde2017panoply} forward POSIX file operations to the
host. \sys cannot adopt this approach because it would expose
security-sensitive metadata, such as file names, file sizes and directory
structures. Instead, it provides complete in-enclave file system
implementations via the Linux \emph{virtual file system}~(VFS)
layer~\cite{vfs}. The VFS layer only requires two host operations for
block-level disk I/O (\code{disk\_read()} and \code{disk\_write()}, see
Table~\ref{tab:hostinterface}).

Applications operate on files through file descriptors as usual, which are
handled by the \emph{ext4} file system implementation of LKL. LKL forwards
block-level I/O requests to a \emph{virtio block device} backend implemented by
\sgxlkllib, which issues \code{disk\_read()} and \code{disk\_write()} calls. On
the host, the reads and writes are made to a single \emph{ext4} disk image
file. The image is mapped into memory by \sysloader. Since the image file has a
fixed size, the read/write operations can be implemented efficiently by
memory-mapping the file and directly accessing the mapped region from within
the enclave.

This approach has three advantages: (i)~it maintains a small host interface
with only 2~disk I/O calls; (ii)~it ensures that individual file accesses are
not visible to the host, which can only observe reads/writes to disk block
offsets; and (iii)~the in-enclave VFS implementation supports different file
systems, such as the \code{/tmp}, \code{/proc} and \code{/dev} in-memory file
systems.

\mypar{TCP/IP network stack} To provide POSIX sockets, \sys uses LKL's TCP/IP
stack for in-enclave packet processing. This: (i)~minimises the host interface
because it only accesses a virtual network device to send/receive Ethernet
frames; (ii)~enables \sys to support any transport protocol (\eg UDP) without
extra host calls; and (iii)~exposes Linux features such as packet encryption
(see~\S\ref{sec:hardening:network}).

To send/receive network traffic, \sysloader sets up a layer-2 TAP device. \sys
implements a corresponding \emph{virtio network device} backend inside the
enclave. To be notified about incoming/outgoing packets, the backend issues a
\code{net\_poll()} request. The return value indicates if the device is ready
for reading/writing packets using \code{net\_read()} and \code{net\_write()}.

\mypar{Memory management} \sys does not interact with the host for memory
allocations/deallocations but a limitation of SGX version~1 is that the enclave
size must be fixed at initialisation time.
\sys therefore pre-allocates enclave memory and provides low-level memory
management primitives inside the enclave. When an enclave is created, it
initially contains \sgxlkllib,
%
%
and an uninitialised heap area. The heap area is exposed through both LKL
functions (\code{mmap()}, \code{mremap()}, \code{munmap()}) and higher-level
\emph{libc} functions (\code{malloc()}, \code{free()}).

\sys supports both variable- and fixed-address anonymous mappings, and tracks
free pages via a heap allocation bitmap. It implements \code{mmap()} by
scanning the bitmap for consecutive free pages large enough for the requested
allocation. To support private file mappings, files are loaded into enclave
because SGX enclaves are bound to a linear address space.

\sys must support changes to page permissions, \eg when loading
libraries. Applications may also modify permissions directly: \eg the Java
Virtual Machine~(JVM) requires executable pages for just-in-time~(JIT)
compilation and changes of permissions for guard pages during garbage
collection. While SGX pages have their own permissions, SGX version~1 requires
permissions to be fixed on enclave creation.


As a workaround, \sys uses an extra \code{mem\_protect()} host call for SGX
version 1 execution. All enclave pages are created with full SGX page
permissions and the actual permissions are set via the host-controlled page
table permissions. Since relying on the host to manage page permissions is a
security risk, SGX version~2 adds the ability to control page permissions from
within the enclave.

\mypar{Thread management} SGX allows multiple host threads to enter an enclave,
The maximum number of \emph{enclave threads}, however, must be specified at
enclave creation time, which prevents dynamic thread creation. In addition, a
one-to-one mapping between enclave and application threads means that the
creation, joining as well as synchronisation of threads requires host OS
support, posing a security risk~\cite{weichbrodt2016asyncshock}.

Therefore, \sys implements user-level threading based on the \emph{lthread}
library~\cite{lthread} and provides synchronisation primitives inside the
enclave. A fixed number of host threads are assigned to enclave threads, which
enter the enclave at startup and only leave when idle. Application threads and
LKL kernel threads are \emph{lthreads}, managed via the standard
\emph{pthreads} interface. \sys implements \emph{futex} calls in-enclave to provide
synchronisation primitives such as mutexes and semaphores.

\mypar{Signals} Applications can register custom signal handlers to handle
exceptions, interrupts or user-defined signals. Some signals (\code{SIGALRM})
are handled entirely within the enclave; others (\code{SIGSEGV}, \code{SIGILL})
are caused by hardware exceptions, exit the enclave and return control to the
host OS. \sys forwards these signals from the host to in-enclave handlers.

During initialisation, \sysloader registers signal handlers for all catchable
signals with the host. All signals are forwarded via
\code{forward\_signal()}, which hides application-specific handlers. \sys
then checks for a corresponding application-registered signal handler and, if
present, delivers the signal. Since application-registered signal handlers
are managed within the enclave, calls such as \code{sigaction()} and
\code{sigprocmask()} are supported without host interaction.

\mypar{Time} Applications, libraries and LKL frequently access time
information. In addition, \sys reads the current time between context switches
to reschedule blocked threads to trigger timer interrupts. Current SGX
implementations, however, do not offer a high-performance in-enclave time
source, and \sys relies on time provided by the host.

Instead of issuing expensive individual host calls, \sys uses the \emph{virtual
  dynamic shared object}~(vDSO)~\cite{vdso} mechanism of the Linux kernel. The
kernel maps a small shared library to the address space outside of the
enclave. On each clock tick, the host kernel updates a shared memory location
with the current time for various clock sources, which are read from within the
enclave when a time-related call is made.

For high precision, the vDSO mechanism requires the \code{RDTSCP} instruction
to adjust for the time passed since the last vDSO update. In SGX version~1,
this instruction is not permitted inside enclaves. The accuracy of
\code{clock\_gettime()} in \sys thus depends on the frequency of vDSO updates;
SGX version~2 does not have this limitation.

\mypar{Illegal instructions} Applications may use the \code{RDTSC} instruction
to read the timestamp counter or \code{CPUID} to obtain CPU features. These
instructions, however, are illegal inside SGX version~1 enclaves.
%
%
\sys catches the resulting \code{SIGILL} exception and emulates the
instructions: \code{RDTSC} is executed outside the enclave, and the result is
forwarded via the \code{forward\_signal()} call; for \code{CPUID}, \sys caches
all \code{CPUID} information during enclave set-up. This eliminates the need
for an extra host call and also hides the \code{CPUID} request by the
application.



\section{Protecting the host interface}
\label{sec:hardening}


\subsection{Protecting disk I/O calls}
\label{sec:hardening:filesystem}

Host disk blocks must have their confidentiality and integrity protected. \sys
uses the Linux \emph{device mapper} subsystem~\cite{devicemapper}, which maps
virtual block devices, such as \sys{}'s virtio backend, to higher-level
devices, and allows I/O data to be transformed. File systems such as
\emph{ext4} use a virtual block device, and data is encrypted and integrity
protected transparently before it reaches the underlying device.

\sys uses different device mapper targets: (i)~\emph{dm-crypt}~\cite{dmcrypt}
provides full-disk encryption using AES in XTS mode with the sector number as
an initialisation vector; (ii)~\emph{dm-verity}~\cite{dmverity} offers
volume-level read-only integrity protection through a Merkle tree of disk block
hashes, with the root node stored in memory; and
(iii)~\emph{dm-integrity}~\cite{dmintegrity} provides key-based block-level
read/write integrity protection. \emph{dm-crypt} and \emph{dm-integrity} can be
combined for full disk encryption with read/write integrity via AES-GCM. For
AES-XTS/AES-GCM, \sys uses hardware AES-NI instructions.

\sys combines the different targets to provide both confidentiality and
integrity for block reads/writes, depending on the security requirements of the
application. For example, for an in-memory key-value store such as
\emph{Memcached}~\cite{memcached}, it is sufficient to protect the integrity of
a read-only disk with \emph{dm-verity}: the Memcached binary stored on disk
must have its integrity protected, but no further sensitive data is stored on
the disk. If the application itself is confidential or other application data
is written to disk, \emph{dm-crypt} can be used with either of the integrity
protection targets. \sys's use of LKL also allows it to support other device
mapper based protection targets such as \emph{dm-x}~\cite{chakraborti2017dmx},
which provides full volume-level read/write integrity and replay
protection. The use of different targets affects I/O performance, as we explore
in~\S\ref{sec:eval:diskio}.

\subsection{Protecting network I/O calls}
\label{sec:hardening:network}

\sys must guarantee the confidentiality and integrity of all network data.
Existing TEE runtime systems either require applications to have built-in
support for network encryption~\cite{baumann2014haven, tsai2017graphene,
  shinde2017panoply} or use TLS~\cite{arnautov2016scone}. Some applications
though do not support TLS (\eg Tensorflow) or use other transport protocols
(\eg UDP-based streaming).

Instead, \sys exploits its TCP/IP network stack inside the enclave, which
enables it to provide transparent low-level network encryption. All data
received and sent via the \code{net\_read()} and \code{net\_write()} host calls
is automatically encrypted, authenticated and integrity-protected. To protect
all network traffic, \sys uses \emph{Wireguard}~\cite{wireguard}, a layer 3 VPN
protocol, currently proposed for inclusion in the Linux
kernel~\cite{inclusion}.

\sys sets up Wireguard at initialisation time and exposes the VPN to the
application through a network interface with its own IP address. An application
binding to this IP address is only reachable by trusted nodes in the VPN. Each
Wireguard \emph{peer} has a public/private key pair, which is bound to a VPN IP
address and an endpoint, an (IP, port)-pair through which the VPN is
accessible. Wireguard uses the asymmetric key pairs to establish ephemeral
symmetric session keys to protect messages using authenticated encryption, and
nonces to prevent replay attacks. In contrast to TLS, which uses certificates,
Wireguard identifies parties through public keys. It does not perform key
distribution---\sys binds keys to enclave identities and supports provisioning
of peers' keys (see~\S\ref{sec:attestation}).

\subsection{Protecting event and time calls}
\label{sec:hardening:protectingremaining}

For the remaining calls in Table~\ref{tab:hostinterface}, \sys must ensure that
an adversary cannot learn confidential data or compromise integrity by
providing invalid data.

\mypar{\code{net\_poll()}}
While the \code{eventmask} reveals if the enclave wants to receive/send
packets, this is already disclosed by the presence of \code{net\_read} and
\code{net\_write} calls. An adversary can return a wrong \code{eventmask}: as a
result, either the \code{net\_read} call fails, which can be handled
transparently, or an invalid packet is read that fails Wireguard's integrity
protection (see~\S\ref{sec:hardening:network}).

\mypar{\code{forward\_signal()}}
\sys must ensure that signals correspond to genuine events with valid signal
descriptions; otherwise an adversary can cause an application signal handler to
execute with invalid signal data. For signals due to hardware exceptions, \sys
ensures that the address lies within the enclave range (\eg \code{SIGSEGV}) or
replaces the address with the current instruction pointer for signals that
refer to a faulting instruction (\eg \code{SIGILL} and \code{SIGFPE}). \sys can
be configured to ignore user-controlled signals (\eg \code{SIGINT}).

\mypar{\code{time\_read()}}
A challenge to integrity is that the time cannot be trusted. An adversary may
return a timestamp that is not monotonically increasing and thus causes an
underflow when an application calculates a timespan. \sys therefore checks for
monotonicity for \code{CLOCK\_MONOTONIC\_*} clock sources.
%
%
Future TEE implementations may provide practical trustworthy time sources,
which \sys could use.

\subsection{Runtime attestation and secret provisioning}
\label{sec:attestation}

\sys must execute securely in an untrusted and potentially malicious
environment. For this, parties must (i)~remotely attest that they execute a
trustworthy version of \sys in an SGX enclave; (ii)~deploy applications
securely, \ie guaranteeing the confidentiality and integrity of application
code; and (iii)~provision applications with secrets such as cryptographic keys
and sensitive application configuration.

The above requirements go beyond attestation in current SGX
SDKs~\cite{intelsdk, openenclave}, which assume that all application code is
compiled into the library, and a single measurement suffices to verify
integrity. They also do not protect application confidentiality, and leave it
to applications to implement custom mechanisms for secret provisioning, which is
cumbersome.

\begin{figure}[tb]
    \centering
    \includegraphics[width=\linewidth]{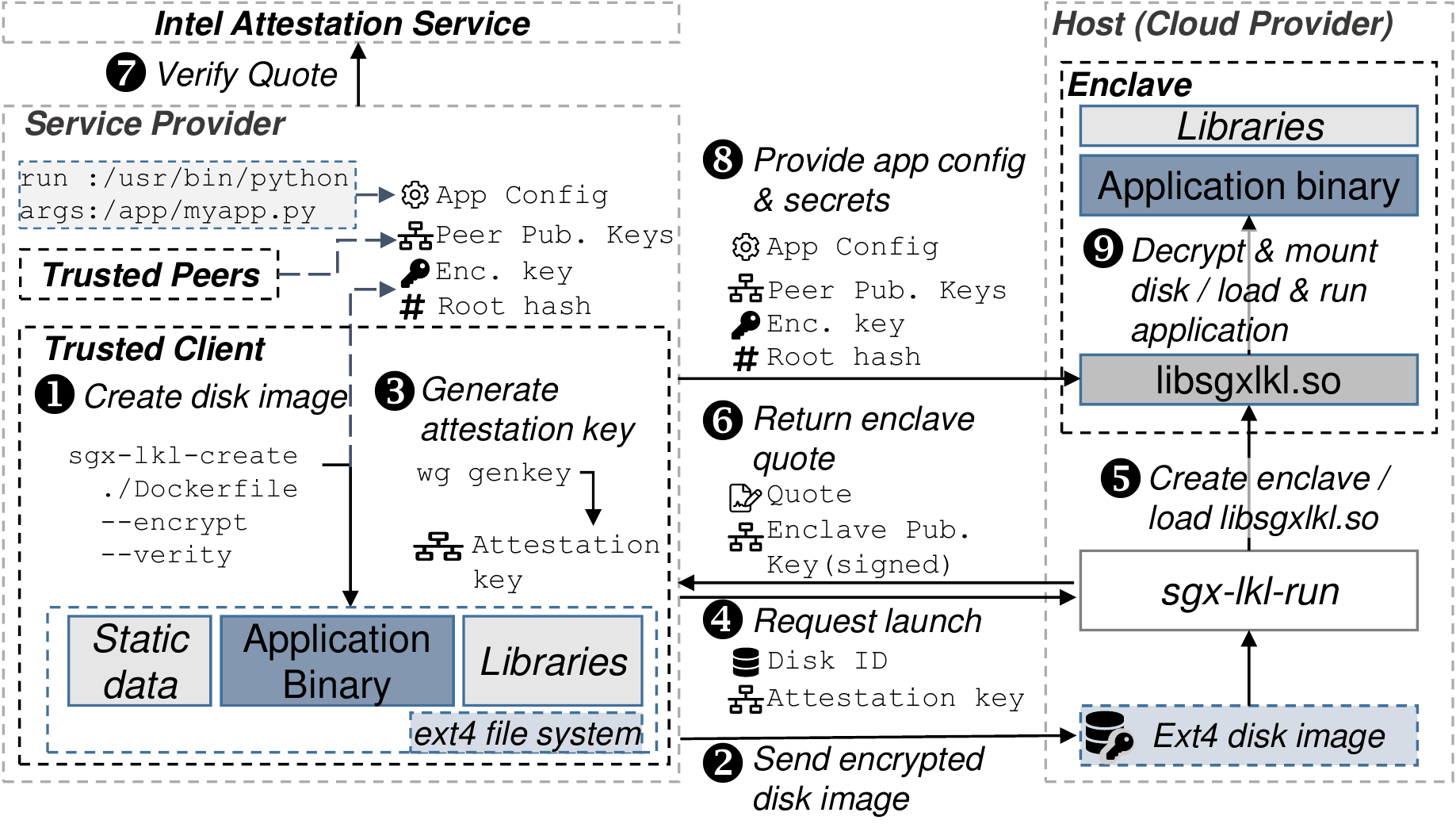}
    \caption{Deployment workflow for \sys}
\label{fig:deployment}
\end{figure}

\sys addresses these issues as part of three phases: (i)~\emph{application
  provisioning}, (ii)~\emph{remote attestation} and (iii)~\emph{secret
  provisioning}.
\F\ref{fig:deployment} shows the deployment workflow, involving three parties:
(i)~a \emph{service provider}~(SP) that wants to deploy an application and has
a trusted client. For a distributed application, this may involve deploying
multiple trusted peers; (ii)~an untrusted host controlled by a \emph{cloud
  provider}~(CP) provides enclaves; and (iii)~the \emph{Intel Attestation
  Service}~(IAS), which allows the SP to verify an enclave measurement.

\mypar{(1)~Application provisioning} In~\ctag{1}, the trusted client creates a
disk image with the application binary and its dependencies, \eg by exporting a
Docker container~\cite{docker}. \sys provides a \code{sgx-lkl-create} tool to
simplify image creation. The tool is based on
\emph{cryptsetup}~\cite{cryptsetup}, which configures \emph{dm-crypt} for disk
encryption/integrity protection. It outputs the encryption key and the root
hash of the Merkle tree for \emph{dm-verity} protection. In~\ctag{2}, the disk
image is sent to the CP. For attestation, the client generates a Wireguard
asymmetric key pair in~\ctag{3}.

\mypar{(2)~Remote attestation} In~\ctag{4}, the public attestation key and the
image disk ID are sent to the cloud host. The host creates an enclave with
\sgxlkllib (\ctag{5}). \sgxlkllib boots LKL, sets up networking and generates
its Wireguard key pair.

Now the enclave can be attested: \sgxlkllib creates a report with an enclave
measurement. SGX allows custom data to be included with the report, which \sys
uses to add the generated public key. The report is signed by a \emph{quoting
  enclave}, and the quote, together with the enclave's public key, is returned
to the SP (\ctag{6}). In~\ctag{7}, the SP sends the quote to the IAS, which
returns a verification report that is checked.

\mypar{(3)~Secret provisioning} After successful attestation, the SP
establishes a secure channel with the enclave through its public
key. \sgxlkllib accepts only one Wireguard peer so that no other party can
communicate with the enclave.

In~\ctag{8}, the SP sends to the enclave: (i)~the disk encryption key; (ii)~the
root hash; (iii)~public keys of other trusted peers; and (iv)~configuration
data, including the executable path and its arguments. In~\ctag{9}, \sgxlkllib
mounts the disk and sets up device mapper for decryption/integrity. It adds the
new Wireguard peers, loads the application, and begins execution.



\section{Making the host interface oblivious}
\label{sec:hardening:oblivious}



As explained in \S\ref{sec:background:host_interfaces}, an adversary can
compromise application confidentiality by observing:
%
%
%
(i)~\emph{frequencies} of calls, \ie their number and time intervals. For
example, there may be more host calls when an application processes data on
disk, thus revealing execution information;
(ii)~\emph{sequences} of calls, \ie the call order and the relationships
between them. For example, an application may always execute reads before
writes, allowing an adversary to make inferences about the execution; and
(iii)~\emph{parameters} of calls. Although \sys encrypts data
blocks~(\S\ref{sec:hardening:filesystem}), the block offsets reveal file
locations. Observing the same offset discloses the data layout.

The main idea for mitigation against these side-channels is to make the host
call interface \emph{oblivious}~\cite{GoldreichO96ORAM}, \ie the sequence and
frequency of calls as well as observable parameters become workload
independent. We focus on the disk~(\S\ref{sec:oblivious:disk}) and
network~(\S\ref{sec:oblivious:network}) I/O host calls and discuss other calls
in \S\ref{sec:oblivious:discussion}.

\subsection{Disk I/O calls}
\label{sec:oblivious:disk}

\sys exposes two disk I/O calls to the host: \code{disk\_read()} and
\code{disk\_write()}. These read and write fixed-size encrypted blocks at the
specified offset (see~\S\ref{sec:hardening:filesystem}). To prevent these calls
from revealing information, \sys employs several techniques: (i)~regarding
frequencies, \sys \emph{discretises} the execution of read/write calls into
fixed time-interval rounds. In each round, it executes a fixed number of calls
in a \emph{batch}, potentially adding indistinguishable \emph{dummy} calls;
(ii)~regarding sequences, \sys makes the order of read/write calls
\emph{deterministic} per batch by issuing them in a predefined order, \eg
always executing reads before writes; and (iii)~regarding parameters, \sys
ensures that all call parameters appear \emph{random}, obscuring patterns. \sys
makes accesses \emph{oblivious} by using seemingly random \code{offset}
parameters. Repeated accesses to the same disk block appear indistinguishable.

\label{sec:oblivious:disk:freq}

\mypar{Hiding disk I/O accesses} sys discretises the execution of all calls
into fixed interval batches. Every $t$~time units, \sys performs a single
\code{disk\_read()} followed by a \code{disk\_write()}. When the LKL filesystem
layer~(see~\S\ref{sec:design:system_support}) issues a disk I/O call, instead
of submitting it directly to the host, it is added to a queue. \sys{}'s
\code{lthread} scheduler checks if enough time has elapsed, and then issues the
next call from the queue.

If there are too many read/write calls in the queue, the remaining calls are
delayed until the next batch. Conversely, if there are not enough read/write
calls from the application, \sys issues \emph{dummy} calls to random blocks in
order to pad that batch. This is done through dummy files in the file system
and reading/writing blocks of those files.

The timing parameter~$t$ must be tuned for good performance because
applications issue system calls at different rates: if $t$ is lower than the
application's call rate, \sys will issue more batches with dummy calls; if $t$
is too high, the application will make slow progress.
In our experiments, we use $t=\mathrm{0.1\unit{ms}}$, which works well across
a range of applications workloads.

\label{sec:oblivious:disk:param}

\mypar{Hiding disk I/O parameters} The call
parameters~(see~\T\ref{tab:hostinterface}) still leak information: (i)~the data
in \code{buf}; (ii)~the disk \code{offset}; and (iii)~the buffer
length~\code{len}. To hide \code{buf}, \sys adopts \emph{probabilistic
  encryption}~\cite{probablistic-enc}, which re-encrypts data before writing
it~(see~\S\ref{sec:hardening:filesystem}). This prevents an adversary from
discerning if written data is old or new.


To hide the \code{offset}, two traces of block accesses must be
indistinguishable. File I/O calls from the application pass through the file
system layer, which may handle them from the in-enclave LKL page cache, or
expose them to the host. The file system layer stores the mapping from logical
to physical blocks. To hide repeated block accesses, \sys must:
(i)~\emph{randomise} the disk layout before application execution, hiding the
initial block mapping; and (ii)~\emph{shuffle} blocks by changing the mapping
in a manner opaque to the adversary.

Existing \emph{oblivious RAM}~(ORAM)~\cite{GoldreichO96ORAM,
  StefanovCCS13PathORAM, wang2015circuit, devadas2016onion,
  sumongkayothin2016recursive, liu2019h} constructions can be used to hide
block accesses. LKL's page cache is well-suited to employ square root
ORAM~\cite{GoldreichO96ORAM}---the page cache is analogous to a ``shelter''. As
it is private, it precludes the need to scan it in its entirety for each access
and allows data to be accessed directly. After a block was added to the page
cache, subsequent accesses to the same block are served directly, until the
page is removed, or flushed to disk. If evicted blocks are requested again, the
adversary would observe this. Therefore, \sys must \emph{obliviously reshuffle}
blocks before continuing execution~\cite{GoldreichO96ORAM}.

An \emph{oblivious shuffle} moves blocks around so that an adversary cannot
correlate the blocks before and after the shuffle. \sys uses a
\emph{k-oblivious shuffle}~\cite{patel2018cacheshuffle}.
This assumes that $k$ of the $n$~source blocks are in the private cache. The
algorithm then sequentially moves the $n$~blocks from the source (either from
the disk or from the cache) to the destination according to a randomly
generated permutation of the blocks.

\newcommand\mycommfont[1]{\footnotesize\textcolor{blue}{#1}}
\SetCommentSty{mycommfont}

\begin{algorithm}[t] 
 \footnotesize\linespread{0.8}
 \SetAlgoLined
 \SetNoFillComment
 \KwData{ \hspace{0.2em} FDs: file descriptors of files to be shuffled \\ \hspace{3.1em} LP: logical to physical block mapping for file system}
 \KwResult{LP\_n: new logical to physical block mapping for file system}
 \For(\tcp*[h]{find file with largest size}){fd in FDs} {\label{alg:max:start}
	file\_blk := num\_blocks(fd)\; 
	max\_blk := max(max\_blk, file\_blk)\; 
	num\_shuff\_blk := num\_shuff\_blk + file\_blk\; \label{alg:max:end}
 } 
 \tcc{create donor files to be used for the swap}
 free\_blk := get\_num\_free\_blocks\_in\_fs()\;\label{alg:freeblock}
 num\_donors := free\_blk / max\_blk\; 
 donor\_files := init\_donors(num\_donors)\;\label{alg:donorfiles} 

 \tcc{generate new random permutation of source blocks}
 src\_blks\_perm := map\_src\_blk\_to\_num(FDs)\;\label{alg:order}
 src\_blk\_new\_perm := FisherYatesShuffle(src\_blk\_perm)\;\label{alg:fisher}  
 \For(\tcp*[h]{get fd \& block no. for id}){id in (0 .. num\_shuff\_blk)} {\label{alg:persource} 
 	fd := fd\_for\_id(src\_blk\_new\_perm, id)\;\label{alg:sourcefd}
	block\_no := blk\_no\_for\_id(src\_blk\_new\_perm, id)\;\label{alg:sourceblockno}
	donor = pick\_rand\_donor(num\_donors)\; 


	\tcc{swap source block with donor block}
	fetch\_source\_block(block\_no)\;\label{alg:unread}
	ioctl(\code{\footnotesize{EXT4\_IOC\_MOVE\_EXT}}, fd, donors\_files[donor], block\_no)\;\label{alg:swap} 
 }
 unlink\_all\_files(donor\_files)\;  
\caption{Oblivious ext4 file system shuffle}\label{alg:shuffle}
\end{algorithm}

Alg.~\ref{alg:shuffle} describes the shuffle algorithm. We assume a list of
files that must be shuffled. To protect a single application workload, this
list can be obtained from a trace of opened files; otherwise all files on the
disk can be considered. We divide the disk blocks into: (i)~blocks that belong
to files that must be shuffled, \ie the source array of size~$n$; and (ii)~a
destination array of $n$~unallocated blocks.

First, the algorithm determines the size~$max\_blk$ of the largest file to be
shuffled~(lines~\ref{alg:max:start}--\ref{alg:max:end}) and the number of free
blocks~(line~\ref{alg:freeblock}). It uses this information to create donor
files~(line~\ref{alg:donorfiles}), each of size~$max\_blk$, to fill up the free
blocks. These donor files are used for swapping blocks with the source
files. Next, the algorithm assigns an ordering to the source file blocks that
need to be shuffled~(line~\ref{alg:order}), mapping the set of blocks to
natural numbers. It uses a \emph{Fisher-Yates shuffle}~\cite{Knuth98a} to
produce a random permutation of the blocks~(line~\ref{alg:fisher}). The
algorithm then performs an iteration per source block to be
shuffled~(line~\ref{alg:persource}). At each iteration, it determines the
source file descriptor and block number that maps to the current index
according to the new
permutation~(lines~\ref{alg:sourcefd}--\ref{alg:sourceblockno}). If the source
block is already in the page cache, an unread source block, if any, is accessed
and brought into the cache~(line~\ref{alg:unread}).

Finally, a random donor file is selected, and the source block is swapped with
the block from the donor file~(line~\ref{alg:swap}). The algorithm is
implemented at the LKL ext4 layer because the shuffled blocks must be
decrypted/encrypted by \emph{dm-crypt}, as discussed in
\S\ref{sec:hardening:filesystem}. The ext4 layer also stores the block mapping,
which needs to be modified when blocks are swapped. \sys uses the
\code{EXT4\_IOC\_MOVE\_EXT} \code{ioctl()} call for this, which exchanges the
blocks belonging to two different files, while also updating the underlying
metadata. Once all the block have been swapped, the dummy files are unlinked,
and the corresponding blocks become unallocated again.

Note that while the creation/unlinking of donor files is deterministic, the
locations of donor file blocks are random and thus indistinguishable from
regular file blocks. Accesses for both types are intermixed and thus also
indistinguishable.


\sys exposes two network I/O calls to the host: \code{net\_read()} and
\code{net\_write()}. These read and write a fixed number of bytes from the
network device (see~\S\ref{sec:hardening:network}). To hide the network
communication pattern, \sys uses fixed-size network packets and adds dummy
packets to generate a constant traffic rate between participating trusted
nodes~\cite{van2015vuvuzela, tyagi2017stadium}.

To achieve this, \sys uses the set of Wireguard peers as the participating
trusted nodes~(see~\S\ref{sec:hardening:network}). For each peer, \sys creates
a thread to send dummy packets with random payloads. These packets are
probabilistically encrypted by Wireguard and indistinguishable from regular
packets.

To ensure a constant traffic rate, \sys uses Linux' traffic shaping
capabilities~\cite{tc}. It sets the queueing discipline~(\emph{qdisc}) for the
the Wireguard interface to \emph{hierarchy token bucket}~(HTB)~\cite{tchtb},
which supports rate limits for network connections. For each peer, \sys adds an
HTB class filter for a per-peer rate limit. To optimise throughput of regular
packets, \sys only sends dummy packets when necessary to reach the rate
limit. It adds a \emph{PRIO} qdisc~\cite{tcprio} below each HTB class, which
prioritises packets with higher priority, and sets dummy packets' priority as
lowest. \sys pads all network packets to the MTU size to obtain fixed-size
packets.


This approach has an overhead proportional to the number of trusted nodes. To
scale to more clients and to hide the set of potential recipients, traffic
could alternatively be routed through a single trusted proxy~\cite{bittau2017prochlo,le2013towards,Chaum88}.


\subsection{Other host calls}
\label{sec:oblivious:discussion}

\myparr{Signals} caused by the application disclose information. For example,
if an application executes an illegal instruction, the resulting signal is
observable. \sys hides what instruction caused the signal but cannot hide the
exception~(see~\S\ref{sec:hardening:protectingremaining}). This is a limitation
of SGX---future implementations may handle exceptions without host involvement.

\mypar{Time} The host kernel updates a vDSO memory region with the current time
(see~\S\ref{sec:design:system_support}). It is read by the enclave, which is
observable by an adversary. However, the \sys scheduler accesses time
frequently on each context switch, and an adversary cannot distinguish between
accesses from \sys and the application, hiding application-specific access
patterns.



\section{Evaluation}
\label{sec:eval}

To understand the performance impact of \sys{}'s design choices, we evaluate it
with real application workloads and micro-benchmarks. All experiments use
machines with Intel Xeon~E3-1280 4-core CPUs with 8\unit{MB} LLC and SGX~1
support, 64\unit{GB} RAM and a 10-Gbps NIC. The machines run Ubuntu Linux~18.04
(kernel~4.15.0-46) with SGX driver~v2.5.

The EPC size is 128\unit{MB}, and around 90\unit{MB} are available to user
applications. Since this limit is too low for some of our workloads but will
increase substantially in the future with new SGX
implementations~\cite{intelepcsizes}, we first evaluate \sys in software-only
mode\footnote{\sys executes the same code in both software and hardware modes.}
to ignore the overhead due to EPC paging.  After that, we consider \sys in
hardware mode to validate how real-world SGX hardware impacts performance. When
using the oblivious host interface, we randomise the disk image and use a large
page cache size.


\subsection{Application performance}
\label{sec:eval:application}

We evaluate the performance of \sys with two data-intensive workloads,
TensorFlow~\cite{tf2016mart} and PARSEC~\cite{parsec08chris}.

\label{sec:eval:tensorflow}

\noindent
We use \textbf{TensorFlow}~(TF)~\cite{tf2016mart} for training and
inference. The models are selected from TF's benchmark
suite~\cite{tfbenchmark}, representing different type of networks, including
small (ResNet-34), large (AlexNet, ResNet-101) and deep and low-dimensional
(ResNet-50). The input datasets are CIFAR10~\cite{cifar} and a subset of
ImageNet~\cite{imagenet}. In \sys, all experiments are executed with and
without disk encryption and oblivious host calls. To avoid SGX paging effects,
we run in software mode with an enclave size of 12.5\unit{GB}.


\begin{figure}[tb]
        \centering
    \includegraphics[width=\columnwidth, trim=0 0 0 0,clip]{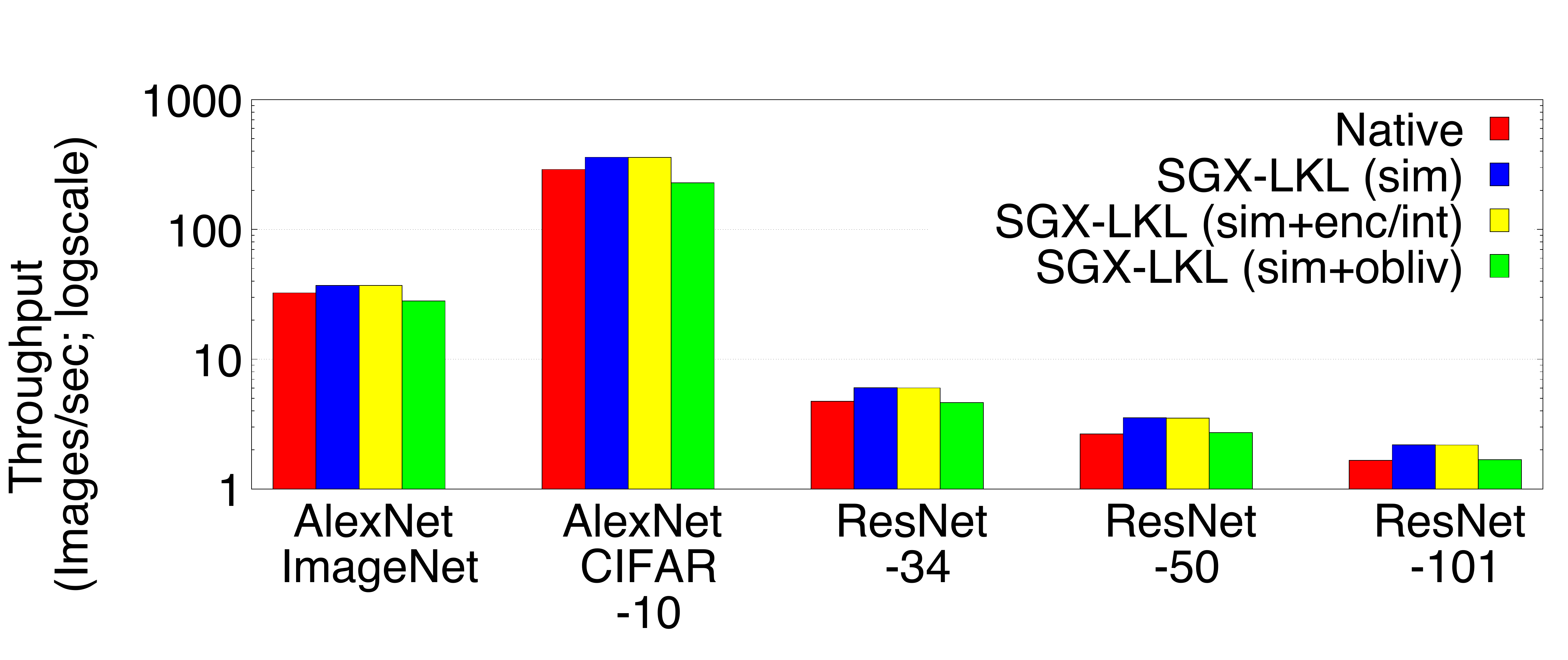}
    \caption{Training throughput with TensorFlow}\label{fig:tf-training-tp}
\end{figure}

\begin{figure}[tb]
    \centering
    \includegraphics[width=\columnwidth, trim=0 0 0 0,clip]{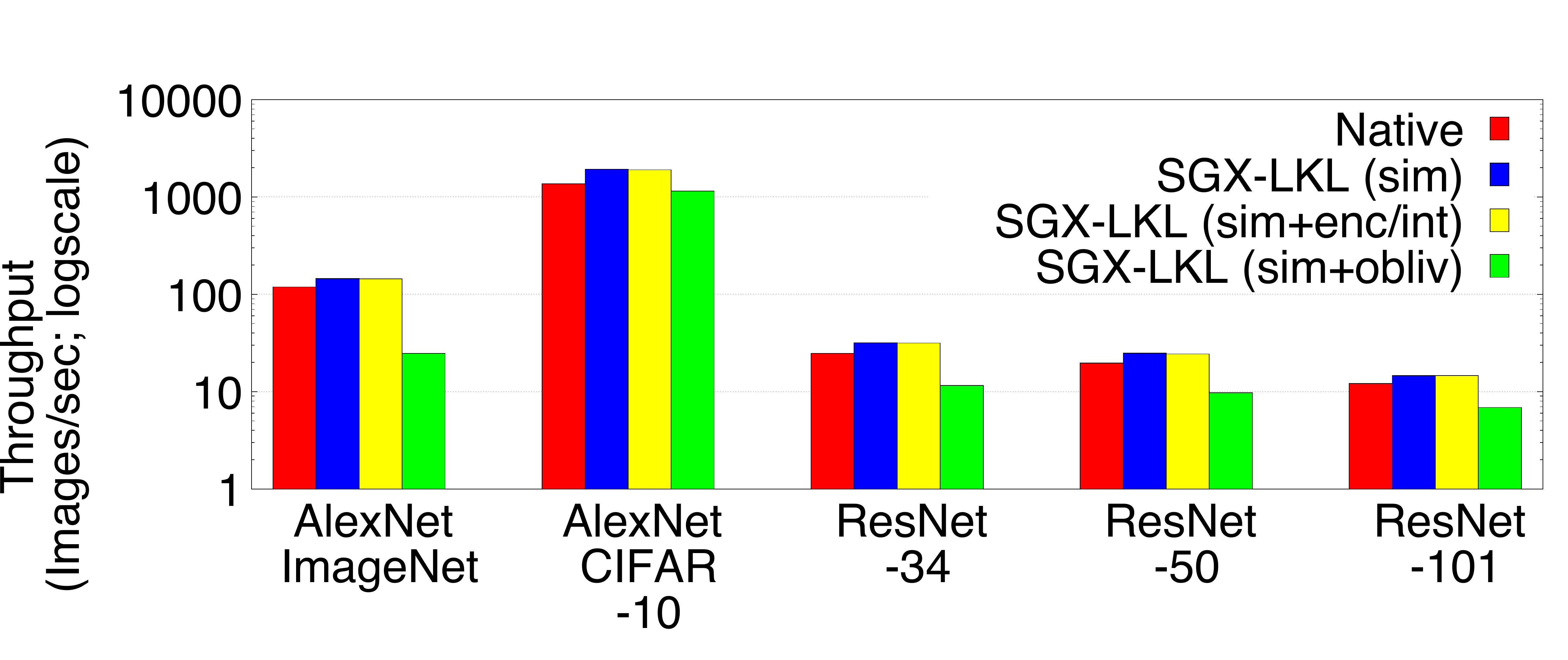}
    \caption{Inference throughput with TensorFlow}
    \label{fig:tf-inference-tp}
\end{figure}

\begin{figure}[tb]
    \centering
    \includegraphics[width=\columnwidth, trim=0 5 0 0,clip]{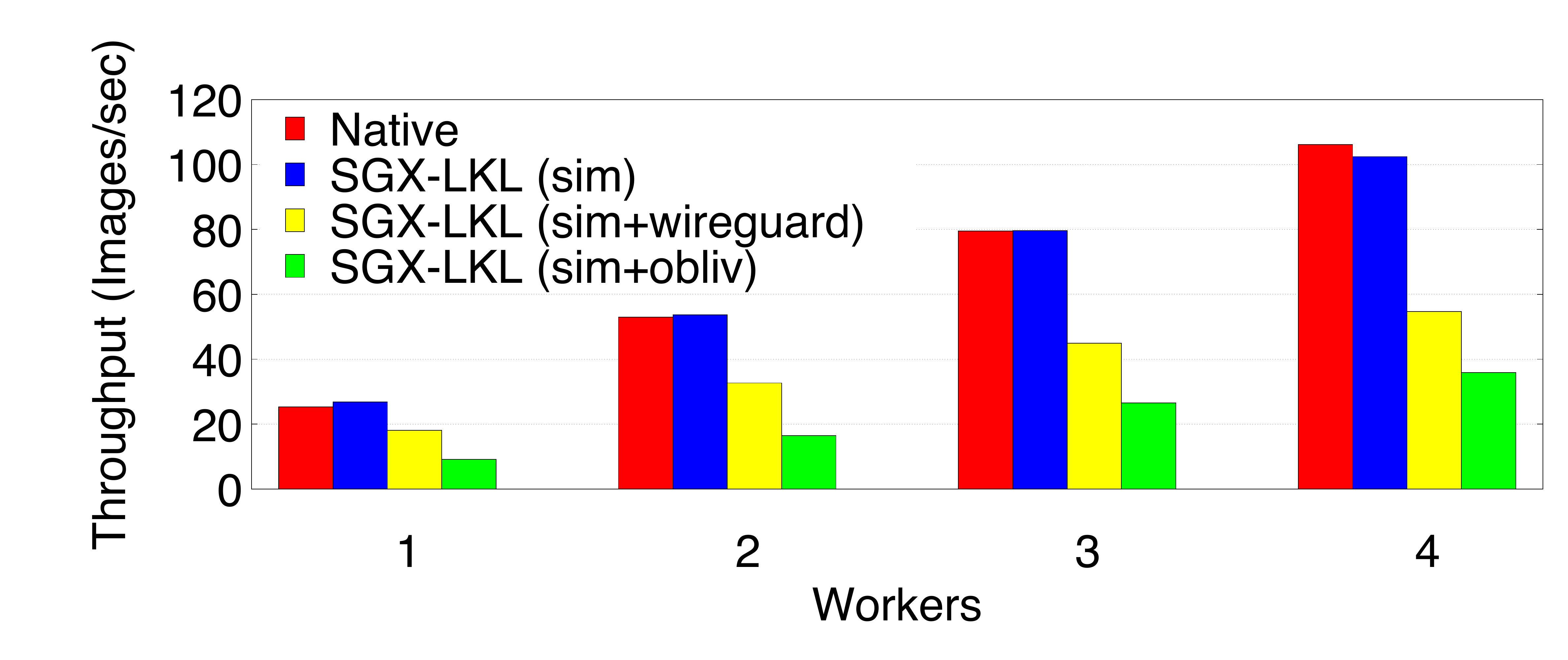}
    \caption{Distributed training throughput with TensorFlow}
    \label{fig:tf-dist}
\end{figure}

In \F\ref{fig:tf-training-tp}, we report the training throughput with \sys in
different modes compared to native execution. \sys outperforms native in
software mode: the training throughput of ResNet-34, ResNet-50 and ResNet-101
is higher than native by 28\%, 33\% and 32\%, respectively.
\sys achieves better performance due to LKL's implementation of system calls as
standard function calls.
The results also show that the overhead of disk encryption and integrity
checking is negligible.

With oblivious host calls with disk encryption, \sys shows a slow-down: the
throughput for AlexNet decreases by 24\%~(ImageNet) and 37\%~(CIFAR10) compared
to software mode, or 14\% and 21\% compared to native execution. The impact of
oblivious host calls on the ResNet models is less: \sys{}'s performance is
almost equal to native. This is due to the different I/O characteristics:
AlexNet processes more images than ResNet per second, thus stressing the
oblivious host calls and requiring more shuffle operations.


In \F\ref{fig:tf-inference-tp}, we also explore inference throughput with \sys
in different modes compared to native.
The results show the same trend as training: the inference throughput is higher
than the native system by 22\% for AlexNet with ImageNet, 42\% for AlexNet with
CIFAR10, 29\% for ResNet-34, 26\% for ResNet 50, and 20\% for ResNet 101.  The
throughput is lower than native with oblivious host calls by 79\% (AlexNet with
ImageNet), 15\% (AlexNet with CIFAR-10), 53\% (ResNet-101), 50\% (ResNet-50),
and 43\% (ResNet-34). We conclude that CPU-intensive networks are less affected
by the oblivious call interface, while I/O-intensive networks are more
affected.

To evaluate performance for distributed applications, we deploy TF with one
parameter server and a varying number of workers, using a batch size of
256.

\F\ref{fig:tf-dist} shows \sys's performance compared to native: without
network encryption, \sys's throughput is comparable to native: with 4~workers,
\sys's throughput is 3\% slower; with the Wireguard VPN, the
throughput decreases to 52\% for 4 workers; and with oblivious network I/O
using a constant rate of 200\unit{Mbps} per peer, the throughput reduces to
34\%. We believe that this is an acceptable security overhead that can be
compensated by scale out.


\label{sec:eval:parsec}

\noindent
To explore the performance of \sys with SGX hardware, we use
five~\textbf{PARSEC} benchmarks~\cite{parsec}, ranging from computer vision
algorithms to machine learning, with different working set sizes. We use four
threads, and the input size is ``simlarge''.

\begin{figure}[t]
    \centering
    \includegraphics[width=\columnwidth, trim=0 45 0 0,clip]{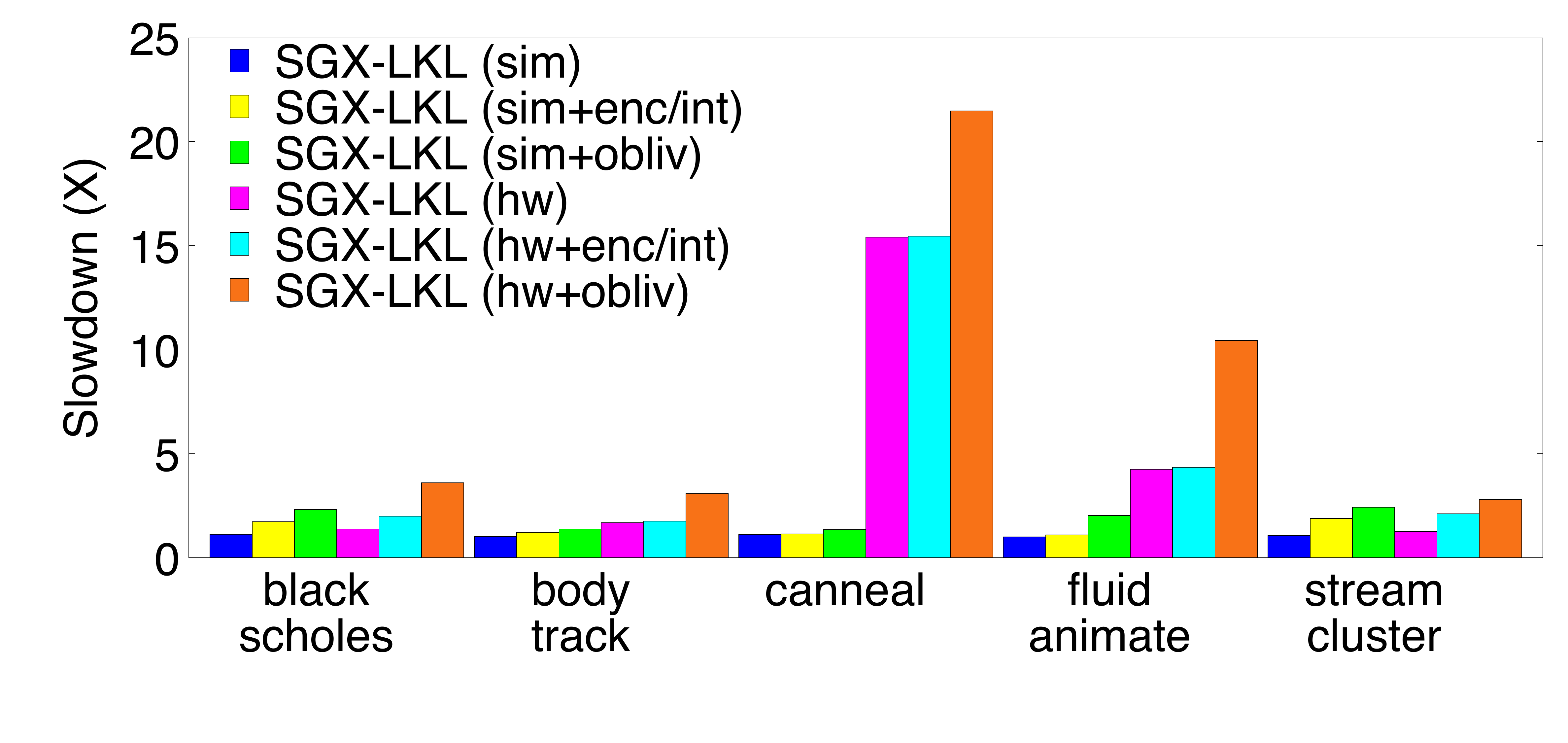}
    \caption{Execution overhead for PARSEC workloads}
    \label{fig:parsec-time}
\end{figure}

\label{sec:eval:diskio}
\F\ref{fig:parsec-time} shows the execution of \sys with different security
guarantees. Without oblivious host calls, \sys in software mode is 1.05$\times$
slower than native. The overhead for \textsf{streamcluster} is 1.07$\times$;
\textsf{fluidanimate} has the same performance as native; for \textsf{canneal}
and \textsf{blackscholes}, the slowdown is around 1.1$\times$ . We note that,
without EPC limits, \sys{}'s performance without oblivious calls is comparable
to native.

In hardware mode, the overhead of workloads with the small to medium working
set sizes is 1.3$\times$ (\textsf{streamcluster}), 1.4$\times$
(\textsf{blackscholes}) and 1.7$\times$ (\textsf{bodytrack}). With large
working sets, \textsf{fluidanimate} and \textsf{canneal} are slower by
4$\times$ and 15$\times$, respectively, which is due to the limited EPC size.

With oblivious host calls, \textsf{streamcluster} shows similar overheads in
software (2.4$\times$) and hardware modes (2.8$\times$). Since
\textsf{streamcluster} does not have input files, no shuffling is required, and
the overhead comes from the dummy calls; for benchmarks with input files, \eg
\textsf{blackscholes}, the overhead in hardware mode is higher (3.6$\times$)
due to shuffling. For \textsf{fluidanimate} and \textsf{canneal}, which have
large memory footprints, the overhead is up to 2$\times$ in software; in
hardware mode, it rises to 10$\times$ and 21$\times$, respectively, due to SGX
paging.

\begin{figure}[tb]
    \centering
    \includegraphics[width=.97\columnwidth, trim = 0 45 0 0, clip]{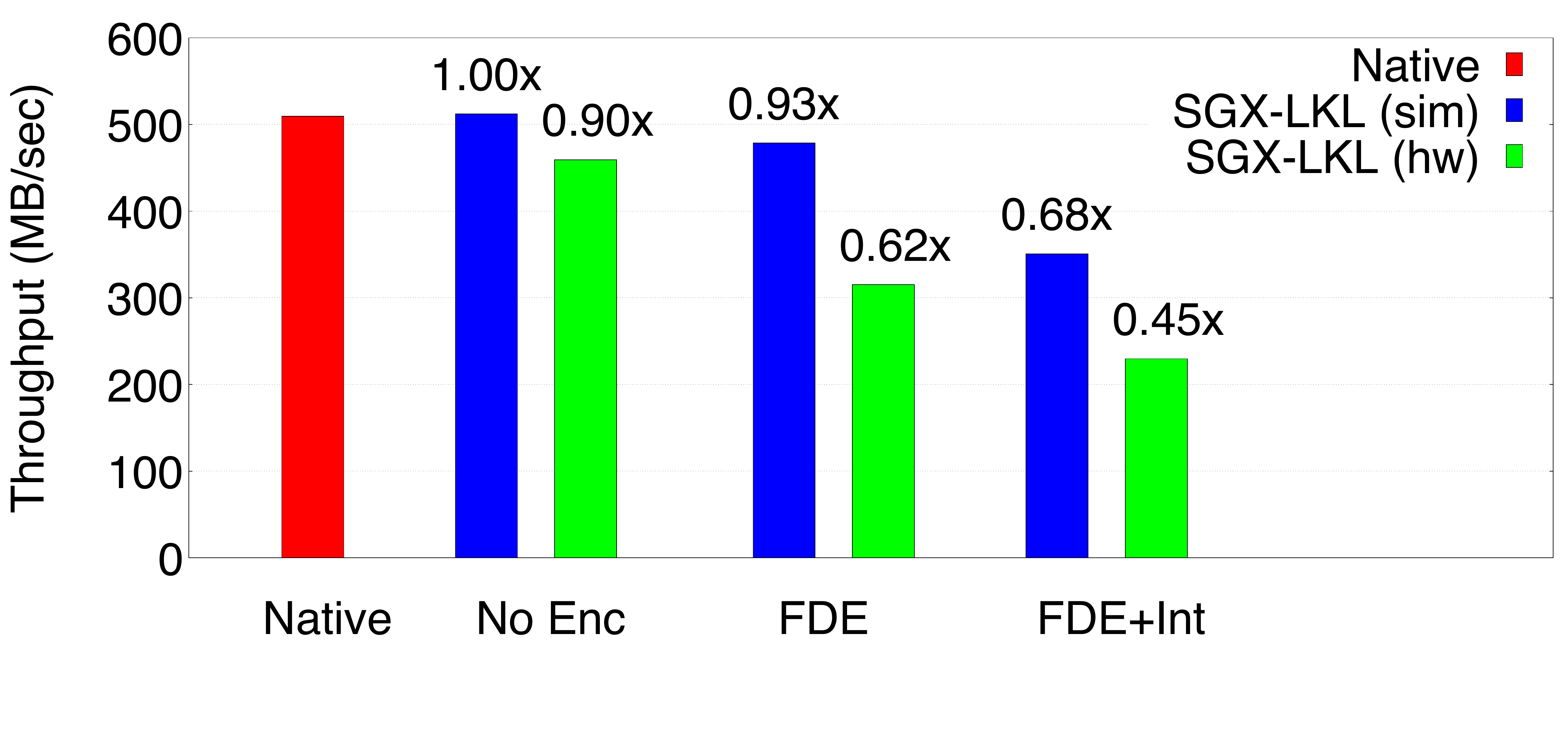}
    \caption{Disk performance with encryption/integrity protection}
    \label{fig:disk}
\end{figure}
\subsection{Disk I/O performance}

We evaluate disk I/O performance by measuring the sequential read throughput
for an uncached 1\unit{GB} file on SSD. We consider: (i)~unencrypted; (ii)~full
disk encryption~(\textsf{FDE}) via \emph{dm-crypt} using AES-XTS; and (iii)~FDE
and integrity protection via \emph{dm-crypt} and \emph{dm-integrity} using
AES-GCM.

\F\ref{fig:disk} shows that \sys achieves near native performance without
encryption or integrity protection, fully saturating the SSD bandwidth of
approx.\ 510\unit{MB/sec}. With \textsf{FDE}, the throughput decreases to
320\unit{MB/sec} (62\%) of native throughput in hardware mode. Enabling
integrity protection further reduces throughput to around
230\unit{MB/sec}~(45\%). This shows the benefit of using x86-specific
cryptographic instructions.

\subsection{Network I/O performance}

 \begin{figure}[tb]
    \centering
    \includegraphics[width=\columnwidth, trim = 0 10 0 0, clip]{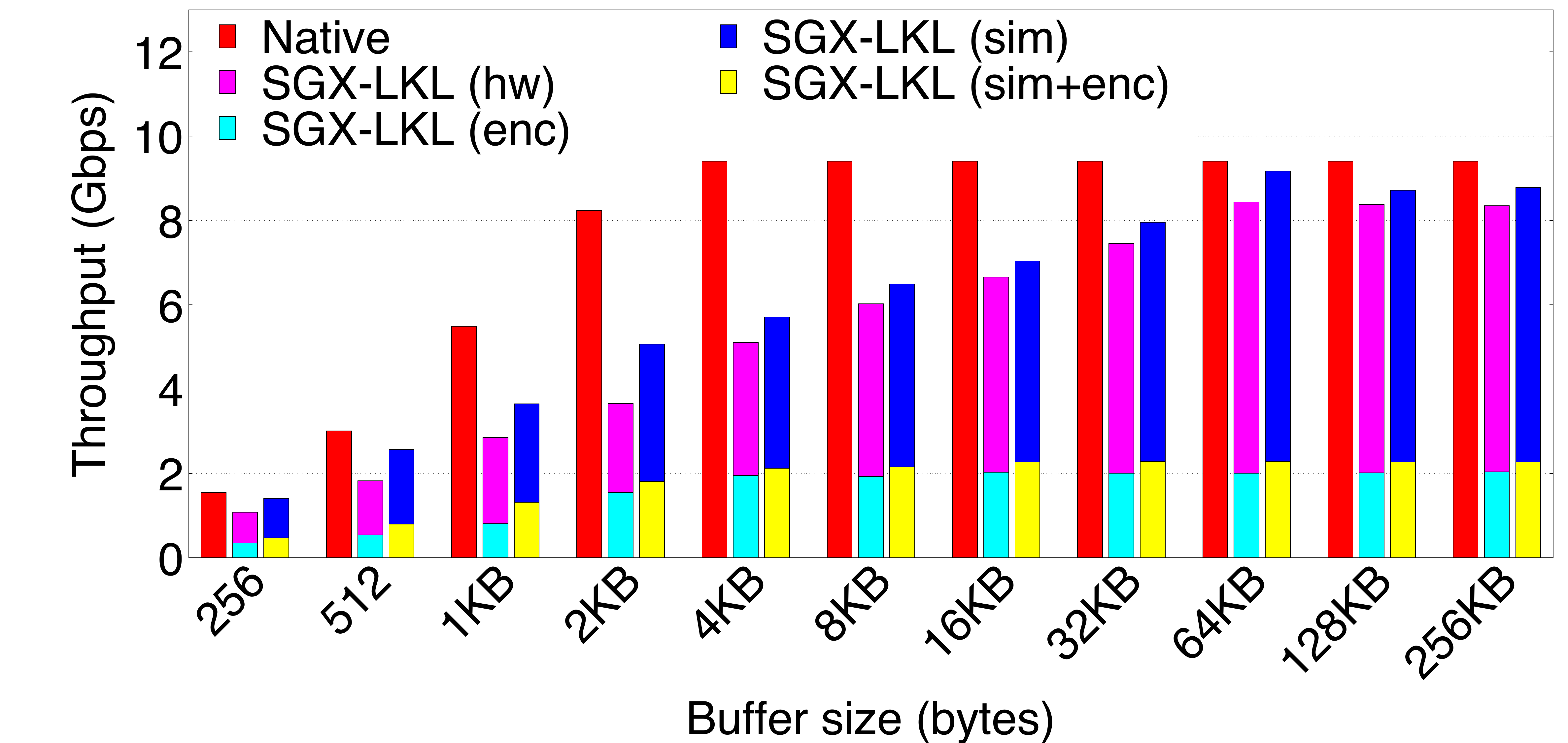}
    \caption{Network I/O throughput for different buffer sizes}
    \label{fig:network}
\end{figure}

We evaluate network performance by measuring throughput for different buffer
sizes with \emph{iperf} v3.1.3~\cite{iperf3}.

\F\ref{fig:network} shows that throughput increases with larger buffer sizes:
for 256\unit{byte} packets, \sys reaches about 0.7$\times$ and 0.9$\times$ of
native throughput in hardware and software modes, respectively; for small
buffers, the throughput is client-bound. Native saturates the network bandwidth
of about 9.4\unit{Gbps} with 4\unit{KB} buffers. Here \sys reaches
5.1\unit{Gbps}~(0.5$\times$) in hardware and 5.7\unit{Gbps}~(0.6$\times$) in
software mode.

Due to its support for TCP segmentation offloading, \sys performs better with
larger buffers: with 64\unit{KB} buffers, it reaches a throughput of
8.4\unit{Gbps} (0.9$\times$) and 8.8\unit{Gbps} (0.9$\times$) for hardware and
software modes, respectively.

With Wireguard and 256\unit{byte} buffers, throughput for hardware and software
mode is 0.44\unit{Gbps} and 0.54\unit{Gbps}, respectively; with buffers larger
than 4\unit{KB}, the throughput increases to 2\unit{Gbps}, and decryption
becomes the bottleneck.



\section{Related Work}
\label{sec:related}

\mypar{TEE runtime systems} As mentioned in
\S\ref{sec:background:host_interfaces}, Haven~\cite{baumann2014haven},
SCONE~\cite{arnautov2016scone}, Graphene-SGX~\cite{tsai2017graphene}, and
Panoply~\cite{shinde2017panoply} are existing TEE runtime systems, but they
have wider host interfaces than \sys and do not support an oblivious host
interface.
%
%
Ryoan~\cite{hunt2018ryoan} uses SGX enclaves to sandbox data processing. It
makes the data flow independent from the content of the input data but does not
make the host interface oblivious.

\myparr{ORAM with TEEs} can conceal access patterns.
ZeroTrace~\cite{sasy2018zerotrace} combines SGX with ORAM to create oblivious
memory primitives. ObliDB~\cite{eskandarian2017oblidb} and
Oblix~\cite{mishra2018oblix} use Path ORAM~\cite{StefanovCCS13PathORAM} with SGX
to hide the access patterns of SQL queries. In Raccoon~\cite{rane2015raccoon},
Path ORAM hides array accesses that depend on secrets. By using ORAM, these
approaches can hide access pattern but hosts can still mount Iago attacks when
returning data blocks to the enclave.

\mypar{Oblivious file systems}
Obliviate~\cite{ahmad2018obliviate} is an oblivious file system for SGX
enclaves. It adapts an ORAM protocol to read/write a data file.
Privatefs~\cite{williams2012privatefs} is an ORAM-based parallel oblivious file
system, which suports concurrent accesses.
Oladi~\cite{crooks2018obladi} is a cloud-based key/value store that supports
oblivious transactions while protecting access patterns from cloud providers.
Bittau \etal\cite{bittau2017prochlo} introduce an oblivious architecture for
monitoring client software behaviour while protecting user privacy.
In constrast to \sys, these approaches do not offer a file system abstraction
with POSIX semantics.

\mypar{Unobservable communication} For message systems, hiding which users
communicate is challenging. Anonymous networks such as
Tor~\cite{dingledine2004tor} are susceptible to traffic analysis
attacks~\cite{hopper2010much, murdoch2005low}. Dining Cryptographers~(DC)
networks~\cite{chaum1988dining,wolinsky2012dissent,sirer2004eluding} provide
stronger guarantees by broadcasting messages to all users, but incur a high
overhead. More recently, Private Information
Retrieval~(PIR)~\cite{chor1995private, corrigan2015riposte} and differential privacy
techniques~\cite{van2015vuvuzela, backes2013anoa, tyagi2017stadium} add noise
to hide metadata. \sys can leverage this to blind communications.



\section{Conclusion}
\label{sec:conclusion}

We described \sys, a TEE runtime system that is designed around a minimal and
oblivious host interface. \sys executes Linux binaries by using the Linux
kernel to provide POSIX abstractions inside of SGX enclaves, and it shuffles
disk blocks to hide access patterns from the host. Ignoring SGX paging effects,
\sys{}'s performance overhead is low, even for complex applications such as
TensorFlow.


{\bibliographystyle{plain}
\bibliography{bibliography}}

\begin{thebibliography}{100}

\bibitem{tf2016mart}
Mart{\'{\i}}n Abadi, Paul Barham, Jianmin Chen, Zhifeng Chen, Andy Davis,
  Jeffrey Dean, Matthieu Devin, Sanjay Ghemawat, Geoffrey Irving, Michael
  Isard, Manjunath Kudlur, Josh Levenberg, Rajat Monga, Sherry Moore,
  Derek~Gordon Murray, Benoit Steiner, Paul~A. Tucker, Vijay Vasudevan, Pete
  Warden, Martin Wicke, Yuan Yu, and Xiaoqiang Zheng.
\newblock Tensorflow: {A} system for large-scale machine learning.
\newblock In {\em 12th {USENIX} Symposium on Operating Systems Design and
  Implementation, {OSDI}}, pages 265--283, 2016.

\bibitem{ahmad2018obliviate}
Adil Ahmad, Kyungtae Kim, Muhammad~Ihsanulhaq Sarfaraz, and Byoungyoung Lee.
\newblock {OBLIVIATE: A Data Oblivious Filesystem for Intel SGX.}
\newblock In {\em Symposium on Network and Distributed System Security (NDSS)},
  2018.

\bibitem{al2012security}
Moshaddique Al~Ameen, Jingwei Liu, and Kyungsup Kwak.
\newblock {Security and Privacy Issues in Wireless Sensor Networks for
  Healthcare Applications}.
\newblock {\em Journal of medical systems}, 36(1):93--101, 2012.

\bibitem{AMDSEV}
AMD.
\newblock {AMD Secure Encrypted Virtualization (SEV)}.
\newblock \url{https://developer.amd.com/sev/}.
\newblock Last accessed: August 7, 2019.

\bibitem{CloudAttacks}
Apriorit.
\newblock {Cloud Computing: A New Vector for Cyber Attacks}.
\newblock
  \url{https://www.apriorit.com/dev-blog/523-cloud-computing-cyber-attacks}.
\newblock Last accessed: August 7, 2019.

\bibitem{Trustzone}
{ARM}.
\newblock {Technologies TrustZone for Cortex-A}.
\newblock
  \url{https://www.arm.com/why-arm/technologies/trustzone-for-cortex-a/tee-reference-documentation}.
\newblock Last accessed: August 7, 2019.

\bibitem{arnautov2016scone}
Sergei Arnautov, Bohdan Trach, Franz Gregor, Thomas Knauth, Andre Martin,
  Christian Priebe, Joshua Lind, Divya Muthukumaran, Dan O'Keeffe, Mark~L.
  Stillwell, David Goltzsche, Dave Eyers, R{\"{u}}diger Kapitza, Peter~R.
  Pietzuch, and Christof Fetzer.
\newblock {SCONE: Secure Linux Containers With Intel SGX}.
\newblock In {\em 12th {USENIX} Symposium on Operating Systems Design and
  Implementation, {OSDI}}, pages 689--703, 2016.

\bibitem{backes2013anoa}
Michael Backes, Aniket Kate, Praveen Manoharan, Sebastian Meiser, and Esfandiar
  Mohammadi.
\newblock {AnoA: A Framework for Analyzing Anonymous Communication Protocols}.
\newblock In {\em 2013 IEEE 26th Computer Security Foundations Symposium},
  pages 163--178, 2013.

\bibitem{baumann2014haven}
Andrew Baumann, Marcus Peinado, and Galen Hunt.
\newblock {Shielding Applications from an Untrusted Cloud with Haven}.
\newblock In {\em Proceedings of the 11th USENIX Conference on Operating
  Systems Design and Implementation}, OSDI, pages 267--283, 2014.

\bibitem{parsec08chris}
Christian Bienia, Sanjeev Kumar, Jaswinder~Pal Singh, and Kai Li.
\newblock {The PARSEC Benchmark Suite: Characterization and Architectural
  Implications}.
\newblock In {\em Proceedings of the 17th International Conference on Parallel
  Architectures and Compilation Techniques}, October 2008.

\bibitem{BiondoCDFS18}
Andrea Biondo, Mauro Conti, Lucas Davi, Tommaso Frassetto, and Ahmad{-}Reza
  Sadeghi.
\newblock The guard's dilemma: Efficient code-reuse attacks against intel
  {SGX}.
\newblock In {\em 27th {USENIX} Security Symposium, {USENIX} Security}, pages
  1213--1227, 2018.

\bibitem{bittau2017prochlo}
Andrea Bittau, {\'U}lfar Erlingsson, Petros Maniatis, Ilya Mironov, Ananth
  Raghunathan, David Lie, Mitch Rudominer, Ushasree Kode, Julien Tinnes, and
  Bernhard Seefeld.
\newblock {Prochlo: Strong Privacy for Analytics in the Crowd}.
\newblock In {\em Proceedings of the 26th Symposium on Operating Systems
  Principles}, pages 441--459, 2017.

\bibitem{vfs}
Daniel~P Bovet and Marco Cesati.
\newblock {Understanding the Linux Kernel. OReilly \& Associates}.
\newblock {\em Inc. October 2000}, 2002.

\bibitem{DR.SGX}
Ferdinand Brasser, Srdjan Capkun, Alexandra Dmitrienko, Tommaso Frassetto, Kari
  Kostiainen, Urs M{\"{u}}ller, and Ahmad{-}Reza Sadeghi.
\newblock {DR.SGX: Hardening SGX Enclaves against Cache Attacks with Data
  Location Randomization}.
\newblock {\em CoRR}, abs/1709.09917, 2017.

\bibitem{BrasserMDKCS17}
Ferdinand Brasser, Urs M{\"{u}}ller, Alexandra Dmitrienko, Kari Kostiainen,
  Srdjan Capkun, and Ahmad{-}Reza Sadeghi.
\newblock Software grand exposure: {SGX} cache attacks are practical.
\newblock In {\em 11th {USENIX} Workshop on Offensive Technologies, {WOOT}},
  2017.

\bibitem{CashGPR15}
David Cash, Paul Grubbs, Jason Perry, and Thomas Ristenpart.
\newblock {Leakage-Abuse Attacks Against Searchable Encryption}.
\newblock In {\em Proceedings of the 22nd {ACM} {SIGSAC} Conference on Computer
  and Communications Security}, pages 668--679, 2015.

\bibitem{chakraborti2017dmx}
Anrin Chakraborti, Bhushan Jain, Jan Kasiak, Tao Zhang, Donald Porter, and Radu
  Sion.
\newblock {Dm-x: Protecting Volume-level Integrity for Cloud Volumes and Local
  Block Devices}.
\newblock In {\em Proceedings of the 8th Asia-Pacific Workshop on Systems},
  APSys, pages 16:1--16:7, 2017.

\bibitem{Chaum88}
David Chaum.
\newblock The dining cryptographers problem: Unconditional sender and recipient
  untraceability.
\newblock {\em J. Cryptology}, 1(1):65--75, 1988.

\bibitem{chaum1988dining}
David Chaum.
\newblock {The Dining Cryptographers Problem: Unconditional Sender and
  Recipient Untraceability}.
\newblock {\em Journal of cryptology}, 1(1):65--75, 1988.

\bibitem{iago_attack}
Stephen Checkoway and Hovav Shacham.
\newblock {Iago Attacks: Why the System Call API is a Bad Untrusted RPC
  Interface}.
\newblock In {\em Proceedings of the Eighteenth International Conference on
  Architectural Support for Programming Languages and Operating Systems},
  ASPLOS, pages 253--264, 2013.

\bibitem{chellappa2002perceived}
Ramnath~K Chellappa and Paul~A Pavlou.
\newblock {Perceived Information Security, Financial Liability and Consumer
  Trust in Electronic Commerce Transactions}.
\newblock {\em Logistics Information Management}, 15(5/6):358--368, 2002.

\bibitem{Hyperspace}
Guoxing Chen, Wenhao Wang, Tianyu Chen, Sanchuan Chen, Yinqian Zhang, XiaoFeng
  Wang, Ten{-}Hwang Lai, and Dongdai Lin.
\newblock {Racing in Hyperspace: Closing Hyper-Threading Side Channels on SGX
  with Contrived Data Races}.
\newblock In {\em 2018 {IEEE} Symposium on Security and Privacy, {SP}}, pages
  178--194, 2018.

\bibitem{chor1995private}
Benny Chor, Oded Goldreich, Eyal Kushilevitz, and Madhu Sudan.
\newblock {Private Information Retrieval}.
\newblock In {\em Proceedings of IEEE 36th Annual Foundations of Computer
  Science}, pages 41--50, 1995.

\bibitem{csacloudthreats}
{Cloud Security Alliance}.
\newblock {Top Threats to Cloud Computing Plus: Industry Insights}.
\newblock
  \url{https://cloudsecurityalliance.org/download/top-threats-cloud-computing-plus-industry-insights}.
\newblock Last accessed: August 7, 2019.

\bibitem{sgx14}
Intel Corp.
\newblock {Software Guard Extensions Programming Reference, Ref. 329298-002US}.
\newblock
  \url{https://software.intel.com/sites/default/files/managed/48/88/329298-002.pdf},
  2014.

\bibitem{corrigan2015riposte}
Henry Corrigan-Gibbs, Dan Boneh, and David Mazi{\`e}res.
\newblock {Riposte: An Anonymous Messaging System Handling Millions of Users}.
\newblock In {\em 2015 IEEE Symposium on Security and Privacy}, pages 321--338,
  2015.

\bibitem{crooks2018obladi}
Natacha Crooks, Matthew Burke, Ethan Cecchetti, Sitar Harel, Rachit Agarwal,
  and Lorenzo Alvisi.
\newblock {Obladi: Oblivious Serializable Transactions in the Cloud}.
\newblock In {\em 13th USENIX Symposium on Operating Systems Design and
  Implementation (OSDI)}, pages 727--743, 2018.

\bibitem{cryptsetup}
{cryptsetup}.
\newblock \url{https://gitlab.com/cryptsetup/cryptsetup}.
\newblock Last accessed: August 16, 2019.

\bibitem{cui2018preserving}
Shujie Cui, Sana Belguith, Ming Zhang, Muhammad~Rizwan Asghar, and Giovanni
  Russello.
\newblock {Preserving Access Pattern Privacy in SGX-Assisted Encrypted Search}.
\newblock In {\em 2018 27th International Conference on Computer Communication
  and Networks (ICCCN)}, pages 1--9. IEEE, 2018.

\bibitem{devadas2016onion}
Srinivas Devadas, Marten van Dijk, Christopher~W Fletcher, Ling Ren, Elaine
  Shi, and Daniel Wichs.
\newblock {Onion ORAM: A Constant Bandwidth Blowup Oblivious RAM}.
\newblock In {\em Theory of Cryptography Conference}, pages 145--174. Springer,
  2016.

\bibitem{devicemapper}
{Linux Kernel device mapper framework}.
\newblock \url{https://www.kernel.org/doc/Documentation/device-mapper}.
\newblock Last accessed: August 7, 2019.

\bibitem{dingledine2004tor}
Roger Dingledine, Nick Mathewson, and Paul Syverson.
\newblock {Tor: The Second-Generation Onion Router}.
\newblock Technical report, Naval Research Lab Washington DC, 2004.

\bibitem{dmcrypt}
{DMCrypt}.
\newblock \url{https://gitlab.com/cryptsetup/cryptsetup/wikis/DMCrypt}.
\newblock Last accessed: August 16, 2019.

\bibitem{dmintegrity}
{DMIntegrity}.
\newblock \url{https://gitlab.com/cryptsetup/cryptsetup/wikis/DMIntegrity}.
\newblock Last accessed: August 16, 2019.

\bibitem{dmverity}
{DMVerity}.
\newblock \url{https://gitlab.com/cryptsetup/cryptsetup/wikis/DMVerity}.
\newblock Last accessed: August 16, 2019.

\bibitem{docker}
Docker.
\newblock {Enterprise Container Platform for High-Velocity Innovation}.
\newblock \url{https://www.docker.com/}.
\newblock Last accessed: August 16, 2019.

\bibitem{duncan2012insider}
Adrian~J Duncan, Sadie Creese, and Michael Goldsmith.
\newblock {Insider Attacks in Cloud Computing}.
\newblock In {\em 2012 IEEE 11th International Conference on Trust, Security
  and Privacy in Computing and Communications}, pages 857--862, 2012.

\bibitem{tee_model}
Jan{-}Erik Ekberg, Kari Kostiainen, and N.~Asokan.
\newblock {Trusted execution environments on mobile devices}.
\newblock In {\em 2013 {ACM} {SIGSAC} Conference on Computer and Communications
  Security, {CCS}}, pages 1497--1498, 2013.

\bibitem{eskandarian2017oblidb}
Saba Eskandarian and Matei Zaharia.
\newblock {ObliDB: Oblivious Query Processing Using Hardware Enclaves}.
\newblock {\em arXiv preprint arXiv:1710.00458}, 2017.

\bibitem{GoldreichO96ORAM}
Oded Goldreich and Rafail Ostrovsky.
\newblock Software protection and simulation on oblivious rams.
\newblock {\em J. {ACM}}, 43(3):431--473, 1996.

\bibitem{probablistic-enc}
Shafi Goldwasser and Silvio Micali.
\newblock {Probabilistic Encryption}.
\newblock {\em Journal of computer and system sciences}, 28(2):270--299, 1984.

\bibitem{GotzfriedESM17}
Johannes G{\"{o}}tzfried, Moritz Eckert, Sebastian Schinzel, and Tilo
  M{\"{u}}ller.
\newblock {Cache Attacks on Intel SGX}.
\newblock In {\em Proceedings of the 10th European Workshop on Systems
  Security, {EUROSEC}}, pages 2:1--2:6, 2017.

\bibitem{graphenesourcecode}
{Graphene Library OS with Intel Registered SGX Support}.
\newblock \url{https://github.com/oscarlab/graphene}.
\newblock Last accessed: August 7, 2019.

\bibitem{gruschka2010attack}
Nils Gruschka and Meiko Jensen.
\newblock {Attack Surfaces: A Taxonomy for Attacks on Cloud Services}.
\newblock In {\em 2010 IEEE 3rd International Conference on Cloud Computing},
  pages 276--279, 2010.

\bibitem{gunawi2014bugs}
Haryadi~S Gunawi, Mingzhe Hao, Tanakorn Leesatapornwongsa, Tiratat
  Patana-anake, Thanh Do, Jeffry Adityatama, Kurnia~J Eliazar, Agung Laksono,
  Jeffrey~F Lukman, Vincentius Martin, et~al.
\newblock {What Bugs Live in the Cloud? A Study of 3000+ Issues in Cloud
  Systems}.
\newblock In {\em Proceedings of the ACM Symposium on Cloud Computing}, pages
  1--14, 2014.

\bibitem{hopper2010much}
Nicholas Hopper, Eugene~Y Vasserman, and Eric Chan-Tin.
\newblock {How Much Anonymity Does Network Latency Leak?}
\newblock {\em ACM Transactions on Information and System Security (TISSEC)},
  13(2):13, 2010.

\bibitem{hunt2018ryoan}
Tyler Hunt, Zhiting Zhu, Yuanzhong Xu, Simon Peter, and Emmett Witchel.
\newblock {Ryoan: A Distributed Sandbox for Untrusted Computation on Secret
  Data}.
\newblock {\em ACM Transactions on Computer Systems (TOCS)}, 35(4):13, 2018.

\bibitem{ibmcloud}
{IBM}.
\newblock {IBM Cloud}.
\newblock \url{https://www.ibm.com/cloud}.
\newblock Last accessed: August 7, 2019.

\bibitem{imagenet}
{imagenet}.
\newblock
  \url{http://download.tensorflow.org/example_images/flower_photos.tgz}.
\newblock Last accessed: August 7, 2019.

\bibitem{IntelSGX}
Intel.
\newblock {Intel Software Guard Extensions}.
\newblock \url{https://software.intel.com/en-us/sgx}.
\newblock Last accessed: August 7, 2019.

\bibitem{intelsdk}
Intel.
\newblock {Intel(R) Software Guard Extensions for Linux* OS}.
\newblock \url{https://github.com/intel/linux-sgx}.
\newblock Last accessed: August 16, 2019.

\bibitem{microcode_updates}
{Intel}.
\newblock {Resources and Response to Side Channel L1TF}.
\newblock
  \url{https://www.intel.com/content/www/us/en/architecture-and-technology/l1tf.html}.
\newblock Last accessed: August 14, 2019.

\bibitem{iperf3}
{IPerf3}.
\newblock \url{https://iperf.fr}.
\newblock Last accessed: August 7, 2019.

\bibitem{intelepcsizes}
Simon Johnson.
\newblock {Keynote: Scaling Towards Confidential Computing}.
\newblock
  \url{https://systex.ibr.cs.tu-bs.de/systex19/slides/systex19-keynote-simon.pdf}.

\bibitem{kim2008trust}
Dan~J Kim, Donald~L Ferrin, and H~Raghav Rao.
\newblock {A Trust-Based Consumer Decision-Making Model in Electronic Commerce:
  The Role of Trust, Perceived Risk, and Their Antecedents}.
\newblock {\em Decision support systems}, 44(2):544--564, 2008.

\bibitem{Knuth98a}
Donald~Ervin Knuth.
\newblock {\em {The art of computer programming, , Volume III, 2nd Edition}}.
\newblock Addison-Wesley, 1998.

\bibitem{cifar}
Alex Krizhevsky, Vinod Nair, and Geoffrey Hinton.
\newblock {The CIFAR-10 dataset}.
\newblock \url{https://www.cs.toronto.edu/~kriz/cifar.html}.
\newblock Last accessed: August 16, 2019.

\bibitem{KrizhevskySH17}
Alex Krizhevsky, Ilya Sutskever, and Geoffrey~E. Hinton.
\newblock {ImageNet Classification with Deep Convolutional Neural Networks}.
\newblock {\em Commun. {ACM}}, 60(6):84--90, 2017.

\bibitem{le2013towards}
Stevens Le~Blond, David Choffnes, Wenxuan Zhou, Peter Druschel, Hitesh Ballani,
  and Paul Francis.
\newblock {Towards Efficient Traffic-Analysis Resistant Anonymity Networks}.
\newblock {\em ACM SIGCOMM Computer Communication Review}, 43(4):303--314,
  2013.

\bibitem{lee2019keystone}
Dayeol Lee, David Kohlbrenner, Shweta Shinde, Dawn Song, and Krste
  Asanovi{\'c}.
\newblock {Keystone: A Framework for Architecting TEEs}.
\newblock {\em arXiv preprint arXiv:1907.10119}, 2019.

\bibitem{lee2017hacking}
Jae{-}Hyuk Lee, Jin~Soo Jang, Yeongjin Jang, Nohyun Kwak, Yeseul Choi, Changho
  Choi, Taesoo Kim, Marcus Peinado, and Brent~ByungHoon Kang.
\newblock {Hacking in Darkness: Return-oriented Programming against Secure
  Enclaves}.
\newblock In {\em 26th {USENIX} Security Symposium, {USENIX} Security}, pages
  523--539, 2017.

\bibitem{liu2019h}
Liang Liu, Rujia Wang, Youtao Zhang, and Jun Yang.
\newblock {H-ORAM: A Cacheable ORAM Interface for Efficient I/O Accesses}.
\newblock In {\em Proceedings of the 56th Annual Design Automation Conference
  2019}, page~33, 2019.

\bibitem{inclusion}
{WireGuard: Secure Network Tunnel}.
\newblock \url{https://lkml.org/lkml/2019/3/22/95}.
\newblock Last accessed: August 16, 2019.

\bibitem{lsds_spectre}
{LSDS}.
\newblock {spectre-attack-sgx}.
\newblock \url{https://github.com/lsds/spectre-attack-sgx}.
\newblock Last accessed: August 7, 2019.

\bibitem{lthread}
{lthread}.
\newblock \url{https://github.com/halayli/lthread}.
\newblock Last accessed: August 16, 2019.

\bibitem{insiderAttacks}
{McAfee}.
\newblock {5 Devious Instances of Insider Threat in the Cloud}.
\newblock
  \url{https://www.skyhighnetworks.com/cloud-security-blog/5-devious-instances-insider-threat-cloud/}.
\newblock Last accessed: August 7, 2019.

\bibitem{memcached}
Memcached.
\newblock {What is Memcached?}
\newblock \url{https://memcached.org/}.
\newblock Last accessed: August 16, 2019.

\bibitem{microsoftconfidential}
{Microsoft}.
\newblock {Azure Confidential Computing}.
\newblock
  \url{https://azure.microsoft.com/en-gb/blog/azure-confidential-computing}.
\newblock Last accessed: August 7, 2019.

\bibitem{openenclave}
Microsoft.
\newblock {Open Enclave SDK}.
\newblock \url{https://openenclave.io/sdk/}.
\newblock Last accessed: August 16, 2019.

\bibitem{mishra2018oblix}
Pratyush Mishra, Rishabh Poddar, Jerry Chen, Alessandro Chiesa, and Raluca~Ada
  Popa.
\newblock {Oblix: An Efficient Oblivious Search Index}.
\newblock In {\em 2018 IEEE Symposium on Security and Privacy (SP)}, pages
  279--296, 2018.

\bibitem{MoghimiIE17}
Ahmad Moghimi, Gorka Irazoqui, and Thomas Eisenbarth.
\newblock {CacheZoom: How SGX Amplifies the Power of Cache Attacks}.
\newblock In {\em Cryptographic Hardware and Embedded Systems - {CHES} 2017 -
  19th International Conference}, pages 69--90. Springer, 2017.

\bibitem{murdoch2005low}
Steven~J Murdoch and George Danezis.
\newblock {Low-Cost Traffic Analysis of Tor}.
\newblock In {\em 2005 IEEE Symposium on Security and Privacy (S\&P')}, pages
  183--195, 2005.

\bibitem{musl}
{Musl}.
\newblock \url{https://www.musl-libc.orgs}.
\newblock Last accessed: August 7, 2019.

\bibitem{lift-shift}
NetApp.
\newblock {What Is a Lift and Shift Cloud Migration?}
\newblock
  \url{https://cloud.netapp.com/blog/what-is-a-lift-and-shift-cloud-migration}.
\newblock Last accessed: August 14, 2019.

\bibitem{Varys}
Oleksii Oleksenko, Bohdan Trach, Robert Krahn, Mark Silberstein, and Christof
  Fetzer.
\newblock {Varys: Protecting SGX Enclaves from Practical Side-Channel Attacks}.
\newblock In {\em 2018 {USENIX} Annual Technical Conference, {USENIX} {ATC}},
  pages 227--240, 2018.

\bibitem{oliver2006healthgear}
Nuria Oliver and Fernando Flores-Mangas.
\newblock {HealthGear: A Real-Time Wearable System for Monitoring and Analyzing
  Physiological Signals}.
\newblock In {\em International Workshop on Wearable and Implantable Body
  Sensor Networks {(BSN)}}, pages 4--pp. IEEE, 2006.

\bibitem{optee}
OP-TEE.
\newblock {Open Portable Trusted Execution Environment}.
\newblock \url{https://www.op-tee.org/}.
\newblock Last accessed: August 7, 2019.

\bibitem{osvik2006cache}
Dag~Arne Osvik, Adi Shamir, and Eran Tromer.
\newblock {Cache Attacks and Countermeasures: The Case of AES}.
\newblock In {\em Cryptographers’ Track at the RSA Conference}, pages 1--20.
  Springer, 2006.

\bibitem{panoplysourcecode}
{Panoply Source Code}.
\newblock \url{https://github.com/shwetasshinde24/Panoply}.
\newblock Last accessed: August 7, 2019.

\bibitem{patel2018cacheshuffle}
Sarvar Patel, Giuseppe Persiano, and Kevin Yeo.
\newblock {CacheShuffle: {A} Family of Oblivious Shuffles}.
\newblock In {\em 45th International Colloquium on Automata, Languages, and
  Programming, {ICALP}}, pages 161:1--161:13, 2018.

\bibitem{Drawbridge}
Donald~E Porter, Silas Boyd-Wickizer, Jon Howell, Reuben Olinsky, and Galen~C
  Hunt.
\newblock {Rethinking the Library OS From the Top Down}.
\newblock In {\em ACM SIGARCH Computer Architecture News}, volume~39, pages
  291--304, 2011.

\bibitem{glibc}
GNU project.
\newblock {The GNU C Library (glibc)}.
\newblock \url{https://www.gnu.org/software/libc/}.
\newblock Last accessed: August 7, 2019.

\bibitem{purdila2010lkl}
Octavian Purdila, Lucian~Adrian Grijincu, and Nicolae Tapus.
\newblock {LKL}: The linux kernel library.
\newblock In {\em 9th RoEduNet IEEE International Conference}, pages 328--333,
  2010.

\bibitem{rane2015raccoon}
Ashay Rane, Calvin Lin, and Mohit Tiwari.
\newblock {Raccoon: Closing Digital Side-Channels Through Obfuscated
  Execution}.
\newblock In {\em 24th USENIX Security Symposium (USENIX Security)}, pages
  431--446, 2015.

\bibitem{readsystemcall}
{Read(2) Linux Programmer's Manual}.
\newblock \url{http://man7.org/linux/man-pages/man2/read.2.html}.
\newblock Last accessed: August 14, 2019.

\bibitem{sabt2015trusted}
Mohamed Sabt, Mohammed Achemlal, and Abdelmadjid Bouabdallah.
\newblock {Trusted Execution Environment: What It Is, and What It Is Not}.
\newblock In {\em 2015 IEEE Trustcom/BigDataSE/Ispa}, pages 57--64, 2015.

\bibitem{sasy2018zerotrace}
Sajin Sasy, Sergey Gorbunov, and Christopher Fletcher.
\newblock {ZeroTrace: Oblivious Memory Primitives From Intel SGX}.
\newblock In {\em Symposium on Network and Distributed System Security
  ({NDSS})}, 2018.

\bibitem{schwarz2019zombieload}
Michael Schwarz, Moritz Lipp, Daniel Moghimi, Jo~Van~Bulck, Julian Stecklina,
  Thomas Prescher, and Daniel Gruss.
\newblock {ZombieLoad: Cross-Privilege-Boundary Data Sampling}.
\newblock {\em arXiv preprint arXiv:1905.05726}, 2019.

\bibitem{multizone}
HEX-Five Security.
\newblock {Introducing MultiZone™ Secure IoT Stack - the first Secure IoT
  Stack for RISC-V }.
\newblock \url{https://hex-five.com/}.
\newblock Last accessed: August 7, 2019.

\bibitem{shinde2017panoply}
Shweta Shinde, DL~Tien, Shruti Tople, and Prateek Saxena.
\newblock {Panoply: Low-TCB Linux Applications With SGX Enclaves}.
\newblock In {\em Proceedings of the Annual Network and Distributed System
  Security Symposium (NDSS)}, page~12, 2017.

\bibitem{sirer2004eluding}
Emin~G{\"u}n Sirer, Sharad Goel, Mark Robson, and Doǧan Engin.
\newblock {Eluding Carnivores: File Sharing With Strong Anonymity}.
\newblock In {\em Proceedings of the 11th Workshop on ACM SIGOPS European
  Workshop}, page~19, 2004.

\bibitem{StefanovCCS13PathORAM}
Emil Stefanov, Marten van Dijk, Elaine Shi, Christopher~W. Fletcher, Ling Ren,
  Xiangyao Yu, and Srinivas Devadas.
\newblock Path {ORAM:} an extremely simple oblivious {RAM} protocol.
\newblock In {\em 2013 {ACM} {SIGSAC} Conference on Computer and Communications
  Security, CCS}, pages 299--310, 2013.

\bibitem{sumongkayothin2016recursive}
Karin Sumongkayothin, Steven Gordon, Atsuko Miyaji, Chunhua Su, and Komwut
  Wipusitwarakun.
\newblock {Recursive M-ORAM: A Matrix ORAM for Clients With Constrained Storage
  Space}.
\newblock In {\em International Conference on Applications and Techniques in
  Information Security}, pages 130--141. Springer, 2016.

\bibitem{ta2006splitting}
Richard Ta{-}Min, Lionel Litty, and David Lie.
\newblock {Splitting Interfaces: Making Trust Between Applications and
  Operating Systems Configurable}.
\newblock In {\em 7th Symposium on Operating Systems Design and Implementation
  {(OSDI})}, pages 279--292, 2006.

\bibitem{tchtb}
{tc-htb manpage}.
\newblock \url{https://linux.die.net/man/8/tc-htb}.
\newblock Last accessed: January 11, 2020.

\bibitem{tcprio}
{tc-htb manpage}.
\newblock \url{https://linux.die.net/man/8/tc-prio}.
\newblock Last accessed: January 11, 2020.

\bibitem{tc}
{tc manpage}.
\newblock \url{http://man7.org/linux/man-pages/man8/tc.8.html}.
\newblock Last accessed: January 11, 2020.

\bibitem{techrepublicdropboxleaks}
{Tech Republic}.
\newblock {Dropbox and Box leak files in security through obscurity nightmare}.
\newblock
  \href{https://www.techrepublic.com/article/dropbox-and-box-leak-files-in-security-through-obscurity-nightmare}.
\newblock Last accessed: August 7, 2019.

\bibitem{tensorflow}
{TensorFlow}.
\newblock \url{https://www.tensorflow.org/}.
\newblock Last accessed: August 7, 2019.

\bibitem{tfbenchmark}
{TensorFlow Benchmarks}.
\newblock \url{https://github.com/tensorflow/benchmarks}.
\newblock Last accessed: August 7, 2019.

\bibitem{tsai2017graphene}
Chia-Che Tsai, Donald~E. Porter, and Mona Vij.
\newblock {Graphene-SGX: A Practical Library OS for Unmodified Applications on
  SGX}.
\newblock In {\em Proceedings of the USENIX Annual Technical Conference (ATC)},
  page~8, 2017.

\bibitem{tyagi2017stadium}
Nirvan Tyagi, Yossi Gilad, Derek Leung, Matei Zaharia, and Nickolai Zeldovich.
\newblock {Stadium: A Distributed Metadata-Private Messaging System}.
\newblock In {\em Proceedings of the 26th Symposium on Operating Systems
  Principles}, pages 423--440, 2017.

\bibitem{parsec}
Princeton University.
\newblock {PARSEC}.
\newblock \url{https://parsec.cs.princeton.edu/}.
\newblock Last accessed: August 7, 2019.

\bibitem{van2018foreshadow}
Jo~Van~Bulck, Marina Minkin, Ofir Weisse, Daniel Genkin, Baris Kasikci, Frank
  Piessens, Mark Silberstein, Thomas~F Wenisch, Yuval Yarom, and Raoul Strackx.
\newblock {Foreshadow: Extracting the Keys to the Intel SGX Kingdom With
  Transient Out-Of-Order Execution}.
\newblock In {\em 27th USENIX Security Symposium (USENIX Security)}, pages
  991--1008, 2018.

\bibitem{vanbulck2019tale}
Jo~Van~Bulck, David Oswald, Eduard Marin, Abdulla Aldoseri, Flavio~D. Garcia,
  and Frank Piessens.
\newblock {A Tale of Two Worlds: Assessing the Vulnerability of Enclave
  Shielding Runtimes}.
\newblock In {\em Proceedings of the 2019 ACM SIGSAC Conference on Computer and
  Communications Security}, CCS, page 1741–1758, 2019.

\bibitem{van2015vuvuzela}
Jelle Van Den~Hooff, David Lazar, Matei Zaharia, and Nickolai Zeldovich.
\newblock {Vuvuzela: Scalable Private Messaging Resistant to Traffic Analysis}.
\newblock In {\em Proceedings of the 25th Symposium on Operating Systems
  Principles}, pages 137--152. ACM, 2015.

\bibitem{vdso}
{vDSO manpage}.
\newblock \url{http://man7.org/linux/man-pages/man7/vdso.7.html}.
\newblock Last accessed: August 7, 2019.

\bibitem{wang2018interface}
Jinwen Wang, Yueqiang Cheng, Qi~Li, and Yong Jiang.
\newblock {Interface-Based Side Channel Attack Against Intel SGX}.
\newblock {\em arXiv preprint arXiv:1811.05378}, 2018.

\bibitem{wang2015circuit}
Xiao Wang, Hubert Chan, and Elaine Shi.
\newblock {Circuit ORAM: On Tightness of the Goldreich-Ostrovsky Lower Bound}.
\newblock In {\em Proceedings of the 22nd ACM SIGSAC Conference on Computer and
  Communications Security}, pages 850--861, 2015.

\bibitem{weichbrodt2016asyncshock}
Nico Weichbrodt, Anil Kurmus, Peter Pietzuch, and R{\"u}diger Kapitza.
\newblock {AsyncShock: Exploiting Synchronisation Bugs in Intel SGX Enclaves}.
\newblock In {\em European Symposium on Research in Computer Security}, pages
  440--457. Springer, 2016.

\bibitem{williams2012privatefs}
Peter Williams, Radu Sion, and Alin Tomescu.
\newblock {Privatefs: A Parallel Oblivious File System}.
\newblock In {\em Proceedings of the 2012 ACM Conference on Computer and
  Communications Security}, pages 977--988, 2012.

\bibitem{wireguard}
{WireGuard}.
\newblock \url{https://www.wireguard.com/}.
\newblock Last accessed: August 7, 2019.

\bibitem{wolinsky2012dissent}
David~Isaac Wolinsky, Henry Corrigan-Gibbs, Bryan Ford, and Aaron Johnson.
\newblock {Dissent in Numbers: Making Strong Anonymity Scale}.
\newblock In {\em Presented as Part of the 10th USENIX Symposium on Operating
  Systems Design and Implementation (OSDI)}, pages 179--182, 2012.

\bibitem{xu2015controlled}
Yuanzhong Xu, Weidong Cui, and Marcus Peinado.
\newblock {Controlled-Channel Attacks: Deterministic Side Channels for
  Untrusted Operating Systems}.
\newblock In {\em Security and Privacy (SP), 2015 IEEE Symposium On}, pages
  640--656, 2015.

\bibitem{zhang2010security}
Rui Zhang and Ling Liu.
\newblock {Security Models and Requirements for Healthcare Application Clouds}.
\newblock In {\em 2010 IEEE 3rd International Conference on Cloud Computing},
  pages 268--275, 2010.

\end{thebibliography}

\end{document}